\documentclass[11pt,a4paper]{article}
\usepackage{jheppub}
\pdfoutput=1

\usepackage[utf8]{inputenc}
\usepackage{amssymb}
\usepackage{amsmath}
\usepackage{amsfonts}
\usepackage{graphicx}
\usepackage{xcolor}
\usepackage{xspace}
\usepackage[normalem]{ulem} 
\usepackage{lscape}
\usepackage{ wasysym }
\usepackage{float}
\usepackage{mathrsfs}
\usepackage{slashed}
\usepackage{multirow,rotating}
\usepackage{siunitx}

\usepackage{hyperref}
\hypersetup{
  colorlinks=true,
  allcolors=darkpurple,
  pdfborder={0 0 0},
  linktocpage=true
}

\definecolor{darkpurple}{rgb}{0.5,0,0.5}
\definecolor{cambridgeblue}{rgb}{0.64, 0.76, 0.68}
\definecolor{darkraspberry}{rgb}{0.53, 0.15, 0.34}

\def\gsim{\raise0.3ex\hbox{$\;>$\kern-0.75em\raise-1.1ex\hbox{$\sim\;$}}}
\def\lsim{\raise0.3ex\hbox{$\;<$\kern-0.75em\raise-1.1ex\hbox{$\sim\;$}}}

\newcommand{\ba}[1]{\begin{eqnarray} \label{(#1)}}
\newcommand{\ea}{\end{eqnarray}}

\def\gsim{\raise0.3ex\hbox{$\;>$\kern-0.75em\raise-1.1ex\hbox{$\sim\;$}}}
\def\lsim{\raise0.3ex\hbox{$\;<$\kern-0.75em\raise-1.1ex\hbox{$\sim\;$}}}

\DeclareMathOperator{\re}{Re}
\DeclareMathOperator{\im}{Im}

\newcommand{\at}[1]{\textcolor{blue}{#1}}

\makeatletter
\g@addto@macro\bfseries{\boldmath}
\newcommand\Label[1]{&\refstepcounter{equation}(\mathrm{\theequation})\ltx@label{#1}&}
\makeatother

\graphicspath{{figs/}}

\allowdisplaybreaks

\preprint{\begin{flushright} IFIC/23-43
 \end{flushright}}	

\title{Heavy neutral leptons from kaons in effective field theory}

\author[a]{Rebeca Beltr\'an,}
\emailAdd{rebeca.beltran@ific.uv.es}
\affiliation[a]{AHEP Group, Instituto de F\'{\i}sica Corpuscular --	CSIC/Universitat de Val{\`e}ncia, Apartado 22085, E--46071 Val{\`e}ncia, Spain}

\author[b]{Julian G\"unther,}
\emailAdd{guenther@physik.uni-bonn.de}
\affiliation[b]{Bethe Center for Theoretical Physics \& Physikalisches Institut der 
	Universit\"at Bonn,\\ Nu{\ss}allee 12, 
	53115 Bonn, Germany}

\author[a]{Martin Hirsch,}
\emailAdd{mahirsch@ific.uv.es}

\author[c]{Arsenii Titov,}
\emailAdd{arsenii.titov@df.unipi.it}
\affiliation[c]{Dipartimento di Fisica ``Enrico Fermi'', 
Università di Pisa and INFN, Sezione di Pisa, \\
Largo Bruno Pontecorvo 3, I--56127 Pisa, Italy}

\author[d,e]{Zeren~Simon~Wang}
\emailAdd{wzs@mx.nthu.edu.tw}
\affiliation[d]{Department of Physics, National Tsing Hua University, Hsinchu 300, Taiwan}
\affiliation[e]{Center for Theory and Computation, National Tsing Hua University, Hsinchu 300, Taiwan}

\abstract{In the framework of the low-energy effective theory containing in addition to the Standard Model fields heavy neutral leptons (HNLs), we compute the decay rates of neutral and charged kaons into HNLs.
We consider both lepton-number-conserving and lepton-number-violating four-fermion operators, taking into account also the contribution of active-heavy neutrino mixing. 
Assuming that the produced HNLs are long-lived, we perform simulations and calculate the sensitivities of future long-lived-particle (LLP) detectors at the high-luminosity LHC as well as the near detector of the Deep Underground Neutrino Experiment (DUNE-ND) to the considered scenario. 
When applicable, we also recast the existing bounds on the minimal mixing case obtained by NA62, T2K, and PS191.
Our findings show that while the future LHC LLP detectors can probe currently allowed parameter space only in certain benchmark scenarios, DUNE-ND should be sensitive to parameter space beyond the current bounds in almost all the benchmark scenarios and for some of the effective operators considered it can even probe new-physics scales in excess of 3000 TeV.
}

\begin{document}
\maketitle

%


\section{Introduction}
\label{sec:intro}

A rich program of searches for long-lived particles (LLPs) is planned at the LHC in the next decades~\cite{Curtin:2018mvb,Alimena:2019zri,Feng:2022inv}.
It envisions the construction of several dedicated experiments sensitive to decay lengths of $\mathcal{O}(1-100)$~m.
Some LLP detectors, namely, FASER~\cite{Feng:2017uoz,FASER:2018eoc} and MoEDAL-MAPP1~\cite{Pinfold:2019nqj,Pinfold:2019zwp}, are already operational, whereas several more experiments are planned for the high-luminosity (HL) phase of the LHC.
These include ANUBIS~\cite{Bauer:2019vqk}, CODEX-b~\cite{Gligorov:2017nwh}, FACET~\cite{Cerci:2021nlb}, FASER2~\cite{Feng:2022inv}, MoEDAL-MAPP2~\cite{Pinfold:2019nqj,Pinfold:2019zwp}, and MATHUSLA~\cite{Curtin:2018mvb,Chou:2016lxi,MATHUSLA:2020uve}.
Besides, the Deep Underground Neutrino Experiment (DUNE)~\cite{DUNE:2020lwj,DUNE:2020ypp,DUNE:2020jqi,DUNE:2021cuw,DUNE:2021mtg} presently under construction will allow searching for light LLPs with its near detector (DUNE-ND)~\cite{DUNE:2021tad} at Fermilab as well.

LLPs are present in many models that can account for unresolved
problems in particle physics and cosmology, such as the mechanism of
neutrino mass generation and the nature of dark matter.  The main
focus of phenomenological studies so far has been on various
renormalizable models, including the Higgs portal, the neutrino
portal, the dark photon portal, as well as on non-renormalizable
models featuring long-lived axion-like particles (ALPs), see
\textit{e.g.}~Refs.~\cite{Feng:2022inv,Antel:2023hkf} for reviews and
references.

However, it is plausible that in addition to the renormalizable
couplings, LLPs may interact with the Standard Model (SM) via
effective, non-renormalizable interactions.  In this case, the SM
viewed as an effective field theory (EFT) has to be extended to
include LLPs and their effective interactions.  One such example is
the $N_R$SMEFT~\cite{delAguila:2008ir,Aparici:2009fh,Bhattacharya:2015vja,
  Liao:2016qyd}, which assumes the existence of heavy neutral leptons
(HNLs), $N$, with masses below or around the electroweak scale, $v$.
Recently, a dictionary of tree-level UV completions of the
dimension-six and dimension-seven operators with $N_R$ has been
provided in Ref.~\cite{Beltran:2023ymm}.

There are two basic ways of HNL production at the LHC: (i) directly
from partonic collisions, and (ii) for $m_N \leq$ (few) GeV, in decays
of mesons, which in turn are copiously produced in $pp$ collisions.
The latter way is also the dominant production channel of the HNLs at the DUNE experiment.
Long-lived HNLs produced in the first way via the four-fermion
operators with two quarks and either two or one $N_R$ have been
studied in Ref.~\cite{Cottin:2021lzz} and Ref.~\cite{Beltran:2021hpq},
respectively.  These two sets of effective operators lead to distinct
phenomenology. The pair-$N_R$ operators may enhance the production
cross section while not contributing to the decay of the lightest HNL.
This makes the far LLP detectors introduced above an ideal place to
probe such effective interactions~\cite{Cottin:2021lzz}.  On the
contrary, single-$N_R$ operators, if large enough to enhance the
HNL production, will also induce HNL decays.  Still, displaced-vertex searches at the local LHC detectors (ATLAS and CMS) are very
efficient to probe this set of effective operators~\cite{Beltran:2021hpq}.
The reach of the LLP detectors to the neutrino dipole operator involving an active neutrino and $N_R$ has been recently estimated in Ref.~\cite{Liu:2023nxi}. 
Instead, Ref.~\cite{Delgado:2022fea} has revisited the LEP limits on the 
dimension-five sterile neutrino dipole operator 
(existing for at least two generations of HNLs, $N_1$ and $N_2$), 
taking into account active-heavy neutrino mixing.

Meson decays into long-lived HNLs triggered by non-renormalizable
interactions have been considered in
Refs.~\cite{DeVries:2020jbs,Beltran:2022ast,Barducci:2022gdv,Barducci:2023hzo,Gunther:2023vmz}.
The EFT appropriate for the description of meson decays is 
$N_R$LEFT~\cite{Bischer:2019ttk,Chala:2020vqp,Li:2020lba,Li:2020wxi},
the low-energy theory where the top quark, the Higgs boson and the
heavy SU$(2)_L$ gauge bosons are not present as dynamical degrees of
freedom.  In Ref.~\cite{DeVries:2020jbs}, the authors have
investigated the reach of the proposed LHC far detectors to HNLs
produced in the decays of $D$- and $B$-mesons via single-$N_R$
operators with a charged lepton, demonstrating that the new physics
scale, $\Lambda$, as high as $100$~TeV could be probed by these
detectors.  The same set of single-$N_R$ operators, but with the
$\tau$ lepton, would lead to HNL production in $\tau$ decays at
Belle~II~\cite{Zhou:2021ylt} (see also Ref.~\cite{Han:2022uho}). 
Pair-$N_R$ operators triggering $D$- and $B$-meson decays have been
thoroughly examined in Ref.~\cite{Beltran:2022ast}.  It has been shown
that for certain operators $\Lambda$ as large as 300~TeV and
active-heavy neutrino mixing squared, $|U_{eN}|^2$, as small as $10^{-15}$
could be tested by MATHUSLA, with the reach of ANUBIS being only a
factor of a few smaller.  In Ref.~\cite{Barducci:2022gdv}, the
sensitivities of FASER2, FACET, ANUBIS, CODEX-b, and MAPP, to the
dimension-five sterile neutrino dipole operator have been estimated.  This
operator leads to two-body decays of vector mesons into $N_1$ and
$N_2$ mediated by a photon, $\gamma$, and a subsequent decay of the
heavier HNL $N_2 \to N_1 \gamma$.
Similarly, the sensitivity reach of FASER2 and FACET to the dipole operator coupling an HNL to the photon has recently been investigated in Ref.~\cite{Barducci:2023hzo} for the HNLs produced from meson decays.
Furthermore, Ref.~\cite{Gunther:2023vmz} studied the sensitivity reach of the DUNE-ND (as well as the LHC far detectors) to the HNLs produced from decays of mesons including kaons, both in the minimal scenario and in the EFT, emphasizing on the feasibility of using the $N_R$SMEFT to describe not only the HNLs but also the lightest neutralinos in the R-parity-violating supersymmetry.

In the present work, we perform state-of-the-art numerical simulations and derive the sensitivities of the future LLP detectors at the LHC as well as the DUNE-ND\footnote{We note that the recently approved experiment SHiP~\cite{SHiP:2015vad,Alekhin:2015byh,SHiP:2021nfo} is expected to have less constraining power than DUNE for LLPs produced in kaon decays, and is hence not considered in this work.} to HNLs produced from neutral and charged kaon decays in the $N_R$LEFT.%
\footnote{Leptonic and semi-leptonic decays of kaons 
in the minimal ``3+1'' case have been 
studied in detail in Ref.~\cite{Abada:2016plb}.
For a systematic study of $K \to \pi \nu \overline{\nu}$ in the 
LEFT (without right-handed neutrinos), see Ref.~\cite{Li:2019fhz}.}
First, we study the scenario in which the HNL production is induced by
pair-$N_R$ operators, while the HNL decay proceeds via
active-heavy neutrino mixing $U_{eN}$.  Second, we investigate the
case in which both HNL production and decay are induced by the same
single-$N_R$ operator structure, but with different quark flavor
indices.  Finally, we consider the situation where the production
proceeds via a single-$N_R$ operator as well as via $U_{eN}$, while
the $N$ decays via mixing only.  We consider both
lepton-number-conserving (LNC) and lepton-number-violating (LNV)
operators and discuss the differences.


HNLs can also mediate meson decays. In particular,  
LNV decays $K^{\mp} \to \pi^\pm \ell^\mp \ell^\mp$ mediated by
light sterile neutrinos have been studied in Ref.~\cite{Zhou:2021lnl} adopting the EFT approach. We also mention Ref.~\cite{Fernandez-Martinez:2023phj}. Here, the authors have derived
limits on EFT operators from HNL searches in kaon decays (among
others).

The remainder of the paper is organized as follows.  In
Sec.~\ref{sec:set-up}, we briefly recap neutral kaon mixing, summarize
the $N_R$LEFT operators of interest, and then discuss the HNL
production from kaon decays.  In Sec.~\ref{sec:experiments}, we
describe various experiments we consider and provide the details of
numerical simulations.  Sec.~\ref{sec:results} contains the derived
sensitivities and the relevant discussion.  Finally, we provide summary and conclusions of our work in Sec.~\ref{sec:summary}.
Additionally, we add two appendices at the end of the paper which explain in detail the computation of the decay widths of the kaons into the HNLs and those of the HNLs into a charged lepton and a pion, in the 
EFT framework.
\section{HNLs from kaons in effective field theory}
\label{sec:set-up}
%
%
\subsection{The neutral kaon system}
\label{sec:kaons}

The neutral kaons $K^0$ ($d\bar{s}$) and $\overline{K^0}$ ($s\bar{d}$)
are flavor eigenstates that can be produced in strong interactions.
Weak interactions cause the mixing between these two neutral states.
If $CP$ were a symmetry of the total hamiltonian $\mathcal{H}$
(including strong, electromagnetic, and weak interactions), $CP$
eigenstates would also be eigenstates of $\mathcal{H}$.  With the
convention (see \textit{e.g.}~Ref.~\cite{Buras:1998raa})
\begin{equation}
 \widehat{CP}|K^0\rangle = - |\overline{K^0}\rangle \,, 
 \qquad 
 \widehat{CP}|\overline{K^0}\rangle = - |K^0\rangle \,,
\end{equation}
we can define the $CP$ eigenstates as:
\begin{equation}
 |K_1\rangle = \frac{1}{\sqrt{2}} \left( |K^0\rangle - |\overline{K^0}\rangle \right) \,, 
 \qquad
 |K_2\rangle = \frac{1}{\sqrt{2}} \left( |K^0\rangle + |\overline{K^0}\rangle \right)  \,,
 \label{eq:K1K2}
\end{equation}
where the former has $CP=+1$ and the latter $CP=-1$.
However, $CP$ is mildly violated by weak interactions and therefore the 
mass eigenstates $|K_S\rangle$ and $|K_L\rangle$, 
characterized by definite lifetimes, are 
different from $|K_1\rangle$ and $|K_2\rangle$:
\begin{equation}
 |K_{S}\rangle = 
 \frac{|K_1\rangle +\varepsilon |K_2\rangle}{\sqrt{1 + |\varepsilon|^2}} \,, 
 \qquad
 |K_{L}\rangle =  
 \frac{|K_2\rangle +\varepsilon |K_1\rangle}{\sqrt{1 + |\varepsilon|^2}} \,, 
\end{equation}
where $\varepsilon$ is the parameter accounting for indirect $CP$
violation in neutral kaon decays.  $K_S$ ($K_L$) denotes the neutral
kaon with the shorter (longer) lifetime.  Experimentally,
$|\varepsilon| \approx
2.23\times10^{-3}$~\cite{ParticleDataGroup:2022pth}, and for the
purposes of this work we can safely neglect it. Thus, in what follows
we assume $|K_S\rangle \approx |K_1\rangle$ and $|K_L\rangle \approx
|K_2\rangle$.

\subsection{Effective field theory with right-handed neutrinos}
\label{sec:NRLEFT}

We will work in the framework of the low-energy effective field theory
extended with right-handed neutrinos, $N_R$, dubbed as $N_R$LEFT, see
\textit{e.g.}~\cite{Bischer:2019ttk,Chala:2020vqp,Li:2020lba,Li:2020wxi}.
We assume $N_R$ to be a Majorana particle and allow for both
lepton-number-conserving (LNC) and lepton-number-violating (LNV)
operators.  Charm and bottom meson decays triggered by 
four-fermion effective operators with $N_R$ have been studied in
detail in Refs.~\cite{DeVries:2020jbs,Beltran:2022ast}.  Here, we are
interested in the decays of kaons induced by four-fermion
interactions. These interactions can be grouped into pair-$N_R$
operators given in table~\ref{tab:opsNN} and single-$N_R$ operators
provided in table~\ref{tab:opseN}.%
\footnote{In what follows, we will not consider single-$N_R$ operators 
with an active neutrino $\nu_L$.}
\begin{table}[t]  
\renewcommand{\arraystretch}{1.3}
\centering
 \begin{tabular}[t]{|c|c|c|}
    \hline
    \multicolumn{3}{|c|}{LNC operators} \\
    \hline
    Name & Structure & $N_\mathrm{pars}$ \\ 
    \hline
    ${\cal O}_{dN}^{V,RR}$ &
    $\left(\overline{d_R}\gamma_{\mu}d_R\right)\left(\overline{N_R}\gamma^{\mu}N_R\right)$ & 
     9 \\
    ${\cal O}_{uN}^{V,RR}$ &
    $\left(\overline{u_R}\gamma_{\mu}u_R\right)\left(\overline{N_R}\gamma^{\mu}N_R\right)$ &
    4 \\
    \hline
    ${\cal O}_{dN}^{V,LR}$ &
    $\left(\overline{d_L}\gamma_{\mu}d_L\right)\left(\overline{N_R}\gamma^{\mu}N_R\right)$ & 
    9 \\
    ${\cal O}_{uN}^{V,LR}$ &
    $\left(\overline{u_L}\gamma_{\mu}u_L\right)\left(\overline{N_R}\gamma^{\mu}N_R\right)$ & 
    4 \\
     \hline
 \end{tabular}
 \hspace{0.5cm}
 \begin{tabular}[t]{|c|c|c|}
   \hline
   \multicolumn{3}{|c|}{LNV operators} \\
   \hline
   Name & Structure & $N_\mathrm{pars}$ \\ 
    \hline
    ${\cal O}_{dN}^{S,RR}$ &
    $\left(\overline{d_L}d_R\right)\left(\overline{N_R^c}N_R\right)$ & 
    18 \\
    %
    %
    ${\cal O}_{uN}^{S,RR}$ &
    $\left(\overline{u_L}u_R\right)\left(\overline{N_R^c}N_R\right)$ &
    8 \\
    %
    \hline
    ${\cal O}_{dN}^{S,LR}$ &
    $\left(\overline{d_R}d_L\right)\left(\overline{N_R^c}N_R\right)$ & 
     18 \\
     ${\cal O}_{uN}^{S,LR}$ &
    $\left(\overline{u_R}u_L\right)\left(\overline{N_R^c}N_R\right)$ & 
     8 \\
     \hline
 \end{tabular}
 \caption{Four-fermion operators in $N_R$LEFT, involving two quarks and two $N_R$'s, assuming one generation of $N_R$. 
LNV operator structures require ``+h.c.''.
In the third column, we provide the number of independent real parameters, 
$N_\mathrm{pars}$, associated with each operator structure.}
 \label{tab:opsNN}
\end{table}
%
%
%
%
\begin{table}[t]  
\renewcommand{\arraystretch}{1.3}
\centering
 \begin{tabular}[t]{|c|c|}
    \hline
    \multicolumn{2}{|c|}{LNC operators} \\
    \hline
    Name & Structure \\
    \hline
   ${\cal O}_{udeN}^{V,RR}$ &
    $\left(\overline{u_R}\gamma_{\mu}d_R\right)\left(\overline{e_R}\gamma^{\mu}N_R\right)$ \\ 
    \hline
    ${\cal O}_{udeN}^{V,LR}$ &
    $\left(\overline{u_L}\gamma_{\mu}d_L\right)\left(\overline{e_R}\gamma^{\mu}N_R\right)$ \\
    \hline
    ${\cal O}_{udeN}^{S,RR}$ &
    $\left(\overline{u_L} d_R\right)\left(\overline{e_L} N_R\right)$ \\ 
    ${\cal O}_{udeN}^{T,RR}$ &
    $\left(\overline{u_L} \sigma_{\mu\nu}d_R\right)\left(\overline{e_L} \sigma^{\mu\nu} N_R\right)$ \\
    \hline 
    ${\cal O}_{udeN}^{S,LR}$ &
    $\left(\overline{u_R} d_L\right)\left(\overline{e_L} N_R\right)$ \\ 
    \hline
 \end{tabular}
 \hspace{0.5cm}
 \begin{tabular}[t]{|c|c|}
    \hline
    \multicolumn{2}{|c|}{LNV operators} \\
    \hline
    Name & Structure \\
    \hline
   ${\cal O}_{udeN}^{V,LL}$ &
    $\left(\overline{u_L}\gamma_{\mu}d_L\right)\left(\overline{e_L}\gamma^{\mu}N_R^c\right)$ \\
    \hline
    ${\cal O}_{udeN}^{V,RL}$ &
    $\left(\overline{u_R}\gamma_{\mu}d_R\right)\left(\overline{e_L}\gamma^{\mu}N_R^c\right)$ \\
    \hline
    ${\cal O}_{udeN}^{S,LL}$ &
    $\left(\overline{u_R} d_L\right)\left(\overline{e_R} N_R^c\right)$ \\ 
    ${\cal O}_{udeN}^{T,LL}$ &
    $\left(\overline{u_R} \sigma_{\mu\nu}d_L\right)\left(\overline{e_R} \sigma^{\mu\nu} N_R^c\right)$ \\ 
    \hline 
    ${\cal O}_{udeN}^{S,RL}$ &
    $\left(\overline{u_L} d_R\right)\left(\overline{e_R} N_R^c\right)$ \\ 
    \hline
 \end{tabular}
 \caption{Four-fermion operators in $N_R$LEFT, 
 involving two quarks, one charged lepton, and one $N_R$. 
 Both LNC and LNV operator structures require ``+h.c.''.
 For one generation of $N_R$, there are 36 independent real parameters associated with each operator structure.}
 \label{tab:opseN}
\end{table}
%
Since the top quark is not in the spectrum of the $N_R$LEFT, 
we have $n_u = 2$ and $n_d = n_e = n_\nu = 3$, 
with $n_f$ denoting the number of generations of a fermion $f$. 
In addition, we assume one generation of $N_R$.

Generically, $N_R$ mixes with the active neutrinos at the renormalizable level. Integrating out the $W$ boson leads to the following 
contribution to the effective Lagrangian:
\begin{equation}
 \mathcal{L}_\mathrm{mix} = - \frac{4 G_F}{\sqrt{2}} V_{ij} U_{\ell N} \left(\overline{u_{iL}} \gamma_\mu d_{jL}\right) \left(\overline{e_{\ell L}} \gamma^\mu N_R^c \right) + \text{h.c.}\,,
 \label{eq:Lmix}
\end{equation}
where $G_F$ is the Fermi constant, $V$ is the CKM quark mixing matrix, and $U_{\ell N}$ is active-heavy neutrino mixing. 
For simplicity, we will assume that the HNL mixes with the electron
neutrino only, and consider the single-$N_R$ operators 
with the first-generation leptons only.%
\footnote{Since the mass of the muon is only smaller than the kaon mass by a factor $\sim 4.7$, 
the branching ratios of the kaon decays triggered by the single-$N_R$ operators with the muon will be phase-space-suppressed with respect to those into an electron and an HNL.}
For charged kaon decays, the relevant CKM matrix element is $V_{us}$.
In what follows, we will separate this contribution 
from the corresponding operator in table~\ref{tab:opseN} by denoting its Wilson coefficient (WC) as
\begin{equation}
 c_\mathrm{mix} = -\frac{4 G_F}{\sqrt{2}} V_{us} U_{\ell N}\,,
 \qquad \text{with} \qquad
 |c_\mathrm{mix}| \approx \frac{0.45\, |U_{\ell N}|}{v^2}\,.
 \label{eq:cmix}
\end{equation}
For a more detailed discussion of $N_R$LEFT, 
including the running of the considered operators 
and their matching to the $N_R$SMEFT operators, see Ref.~\cite{Beltran:2022ast}.

\subsection{HNL production in kaon decays}
\label{sec:production}
The pair-$N_R$ operators $\mathcal{O}_{dN}$ from table~\ref{tab:opsNN} 
involving $d$ and $s$ quarks 
trigger the following two- and three-body decays:
$K_{S/L} \to N N$, $K_{S/L} \to \pi^0 N N$ and $K^\pm \to \pi^\pm N N$. 
The single-$N_R$ operators $\mathcal{O}_{udeN}$ from table~\ref{tab:opseN} containing $u$ and $s$ quarks induce
$K^\pm \to e^\pm N$, $K^\pm \to \pi^0 e^\pm N$ and 
$K_{S/L} \to \pi^\pm e^\mp N$. 
In appendix~\ref{app:kaondecays}, we provide the formulae for the two-body
decay widths and the three-body decay amplitudes. 
The three-body decay widths are then computed according to 
the procedure explained in Refs.~\cite{DeVries:2020jbs,Beltran:2022ast}.

If the products of HNL decays are not detected, the kaon decay modes we are interested in contribute to $K_{S/L} \to \text{inv.}$, $K_{S/L} \to \pi^0 \nu \overline\nu$, $K^\pm \to \pi^\pm \nu \overline\nu$, and
$K^+ \to e^+ \nu_e$, $K^+ \to \pi^0 e^+ \nu_e$, $K_{S/L} \to \pi^\pm e^\mp \nu_e$. 
The branching ratios of the decays into final states with at least one charged particle have been measured, 
whereas for $K_L \to \pi^0 \nu \overline\nu$ a stringent upper limit on the branching ratio has been obtained. 
We summarize the current experimental results in table~\ref{tab:Kdecays}.
\begin{table}[t]
\renewcommand{\arraystretch}{1.2}
\centering
 \begin{tabular}[t]{| l | c |}
 \hline
 Decay & Branching ratio \\
 \hline
 \hline
 $K_L \to \pi^0 \nu \overline{\nu}$ & $<3.0 \times 10^{-9}$ at 90\% C.L. \\
 $K^+ \to \pi^+ \nu \overline{\nu}$ & $\left(1.14^{+0.40}_{-0.33}\right) \times 10^{-10}$ \\
 \hline
 \end{tabular}
 \hspace{0.4cm}
 \begin{tabular}[t]{| l | c |}
 \hline
 Decay & Branching ratio \\
 \hline
 \hline
 $K^+ \to e^+ \nu_e$ & $\left(1.582 \pm 0.007\right) \times 10^{-5}$ \\
 $K^+ \to \pi^0 e^+ \nu_e$ & $\left(5.07 \pm 0.04\right) \times 10^{-2}$ \\
 $K_S \to \pi^\pm e^\mp \nu_e$ & $\left(7.04 \pm 0.08\right) \times 10^{-4}$ \\
 $K_L \to \pi^\pm e^\mp \nu_e$ & $\left(40.55 \pm 0.11\right) \times 10^{-2}$ \\
 \hline
 \end{tabular}
 \caption{Branching ratios of semi-invisible kaon decays~\cite{ParticleDataGroup:2022pth}. 
 For $K_{S/L} \to \pi^\pm e^\mp \nu_e$, the values are for the sum 
 of particle and antiparticle states indicated.}
 \label{tab:Kdecays}
\end{table}
%
We will take them into account when deriving 
the limits on the WCs of the pair-$N_R$ 
and single-$N_R$ operators. 
In the case of a measured branching ratio, we will require the new contribution not to exceed twice the experimental error, whereas for 
$K_L \to \pi^0 N N$, we will demand that its branching ratio
is smaller than $3.0 \times 10^{-9}$ 
in accordance with the current upper bound on the branching ratio of 
$K_L\to \pi^0 \nu \bar{\nu}$~\cite{ParticleDataGroup:2022pth}.

\begin{figure}[t]
 %
 \includegraphics[width=0.49\textwidth]{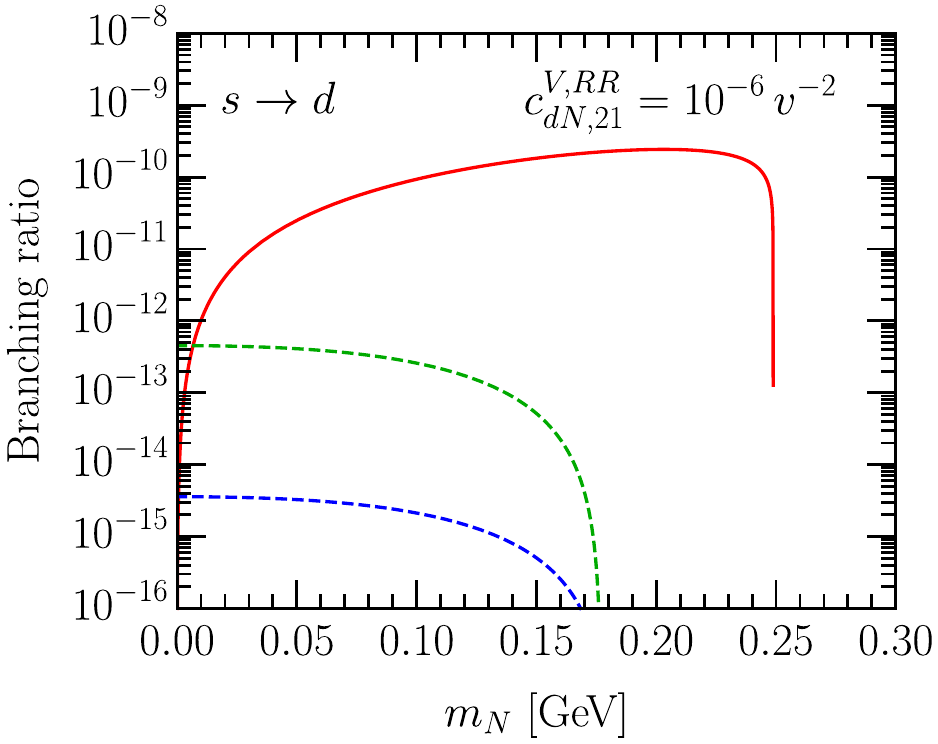}
 \includegraphics[width=0.49\textwidth]{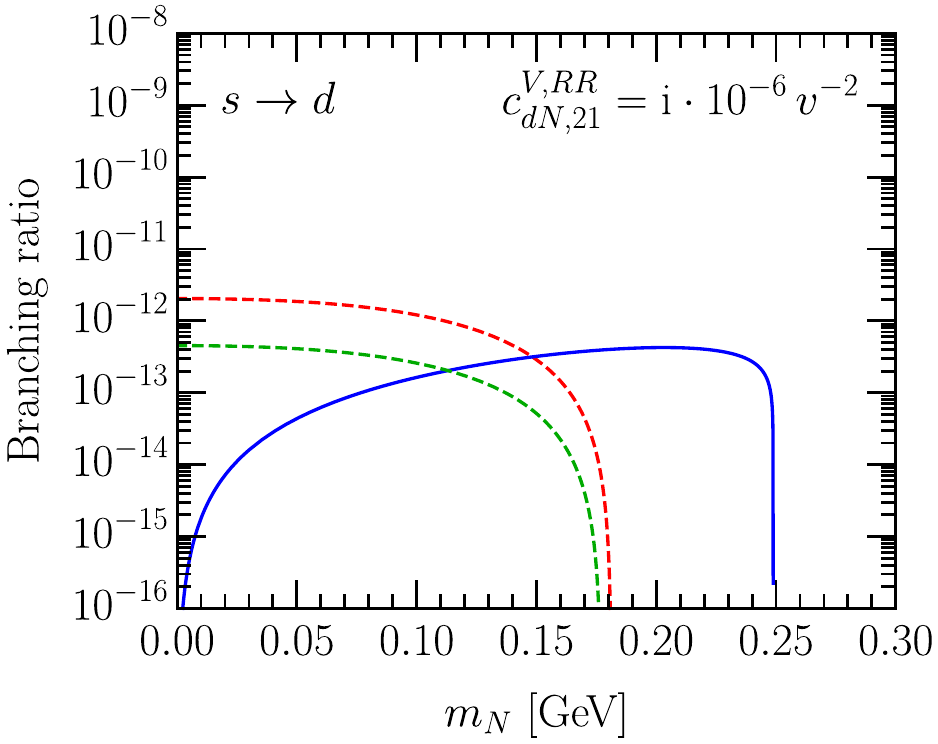}\\
 \includegraphics[width=0.49\textwidth]{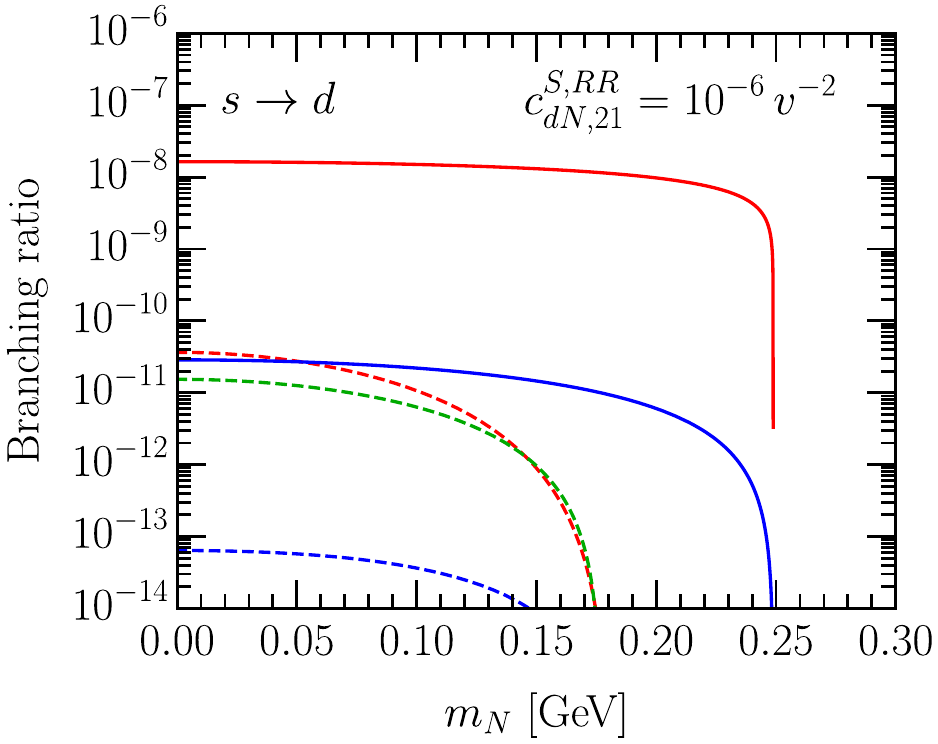}
 \includegraphics[width=0.49\textwidth]{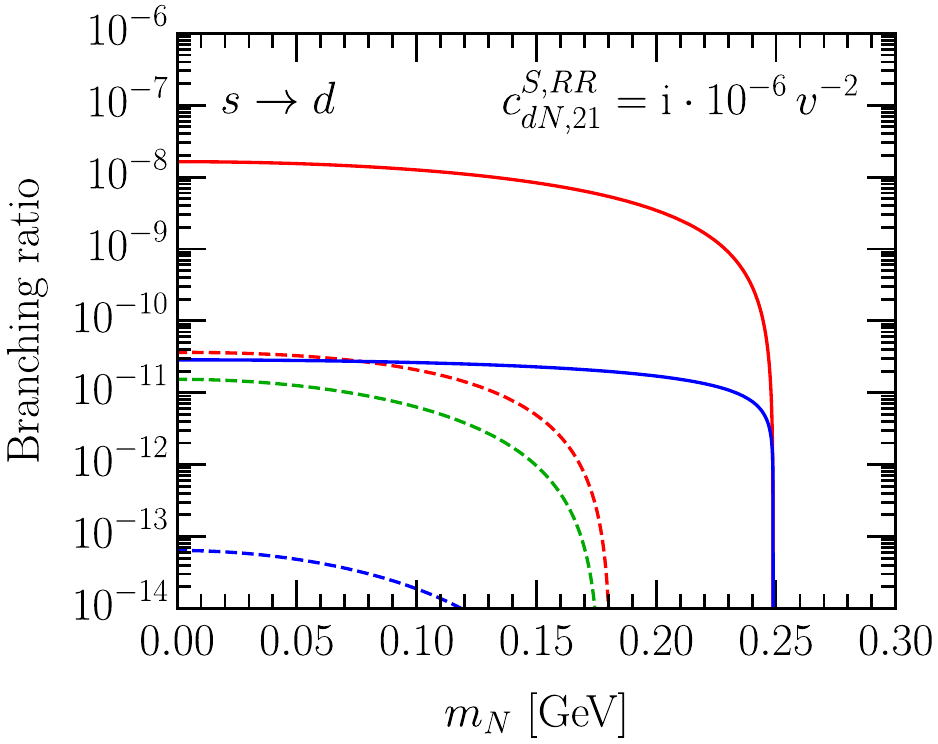}\\[0.2cm]
 \hspace*{1.8cm}
 \includegraphics[width=0.8\textwidth]{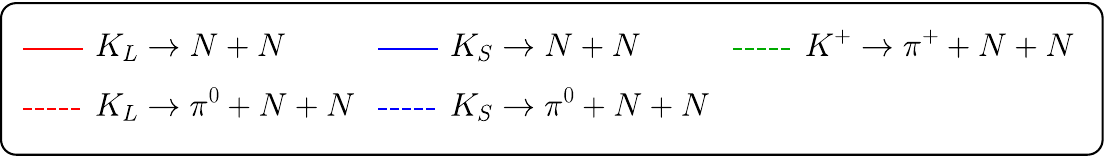}
 \caption{Branching ratios of kaon decays triggered by 
 the LNC (top) and LNV (bottom) pair-$N_R$ operators. 
 In the left (right) panel, the corresponding WC
 is purely real (imaginary).}
 \label{fig:BRsNN}
\end{figure}

In figure~\ref{fig:BRsNN}, we display the branching ratios of kaon decays triggered by the LNC operator $\mathcal{O}_{dN,21}^{V,RR}$ 
and by the LNV operator $\mathcal{O}_{dN,21}^{S,RR}$. 
In each case, we assume that the corresponding WC
is either real (left panel) or purely imaginary (right panel). 
For the LNC operator, a real (purely imaginary) WC does not allow 
for $K_S \to NN$ and $K_L \to \pi^0 NN$ 
($K_L \to NN$ and $K_S \to \pi^0 NN$), 
as can be inferred from eqs.~\eqref{eq:GammaKSNN} and \eqref{eq:GammaKLpi0NN} 
(eqs.~\eqref{eq:GammaKLNN} and \eqref{eq:GammaKSpi0NN}).%
\footnote{We note that this is strictly true only in the limit $\varepsilon \to 0$.} 
For small $m_N$, the allowed two-body decay is suppressed, 
since the corresponding decay width scales with $m_N^2$, 
see eqs.~\eqref{eq:GammaKSNN} and \eqref{eq:GammaKLNN}.
In this figure, we fix the absolute value of the operator coefficient to 
$10^{-6} v^{-2}$ in order to comply with the measurements 
reported in the left panel of table~\ref{tab:Kdecays}, 
in particular, with that of $K^+ \to \pi^+ \nu \overline{\nu}$. 
Switching on $\mathcal{O}_{dN,21}^{V,LR}$ and 
$\mathcal{O}_{dN,21}^{S,LR}$ (one at a time) would lead 
to the same results as for $\mathcal{O}_{dN,21}^{V,RR}$ and 
$\mathcal{O}_{dN,21}^{S,RR}$, respectively.

\begin{figure}[t]
 %
 \includegraphics[width=0.49\textwidth]{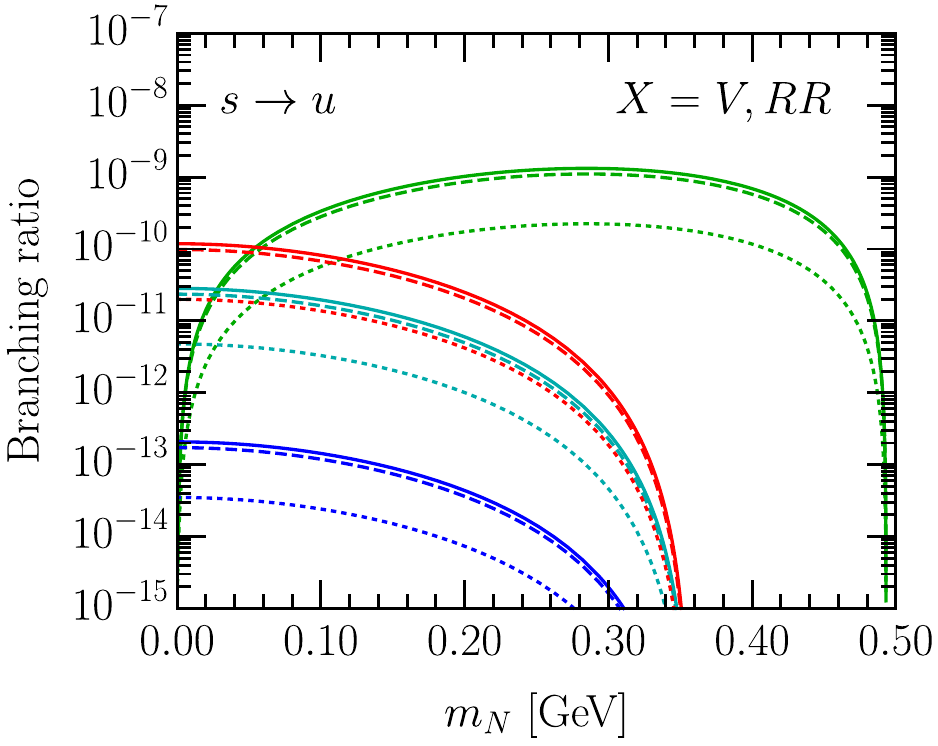}
 \includegraphics[width=0.49\textwidth]{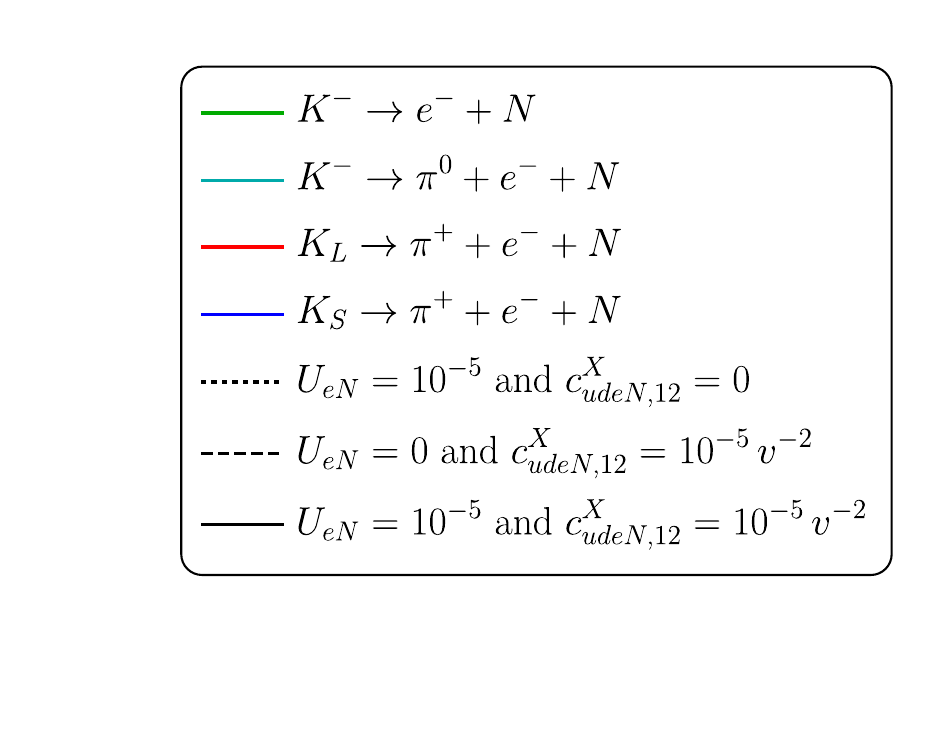}\\
 \includegraphics[width=0.49\textwidth]{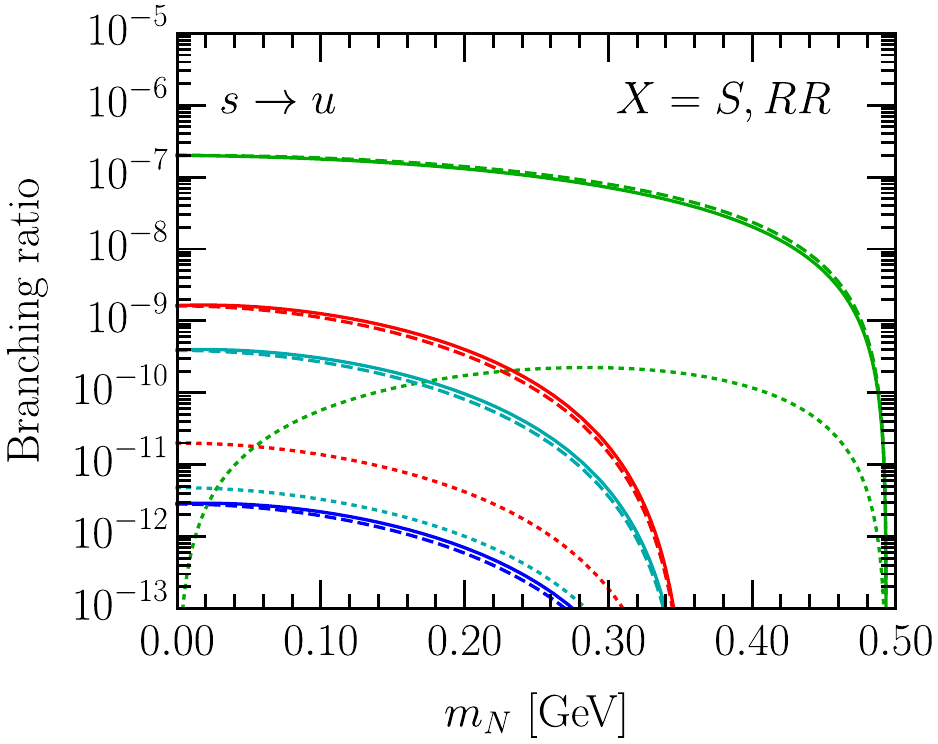} 
 \includegraphics[width=0.49\textwidth]{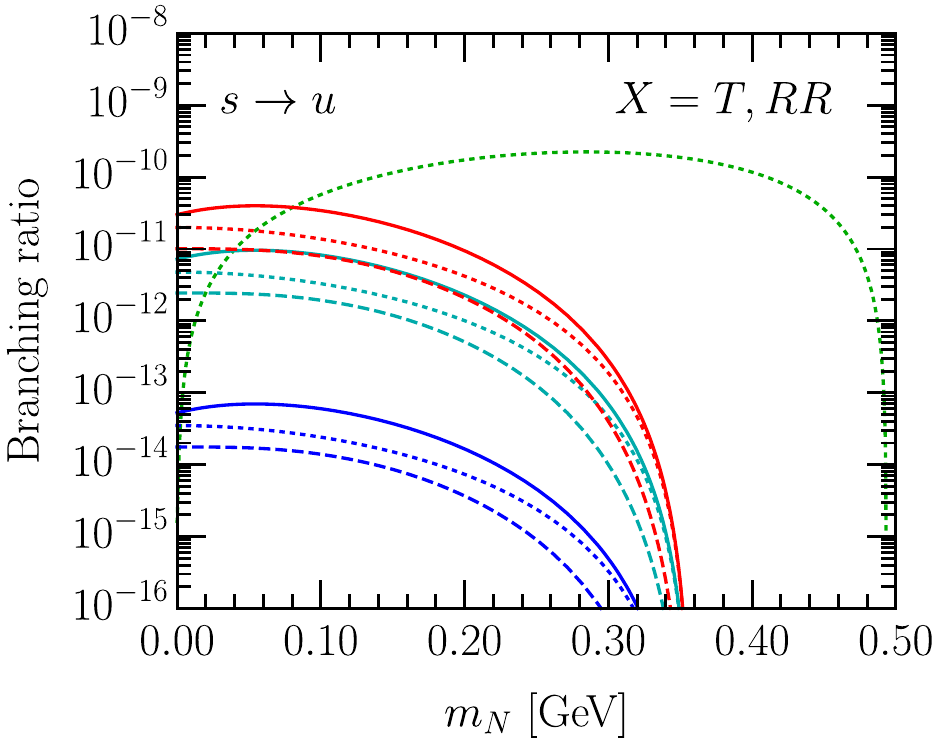}
 \caption{Branching ratios of kaon decays triggered by 
 the LNC single-$N_R$ operators with electron, 
 as well as by active-heavy mixing $U_{eN}$.}
 \label{fig:BRseNLNC}
\end{figure}

In figure~\ref{fig:BRseNLNC}, we plot the branching ratios 
of kaon decays induced by the LNC operators 
$\mathcal{O}_{udeN,12}^{V,RR}$, 
$\mathcal{O}_{udeN,12}^{S,RR}$, and 
$\mathcal{O}_{udeN,12}^{T,RR}$,
with respective WC $c_\mathcal{O}$, 
as well as by active-heavy neutrino mixing $U_{eN}$. 
We show three representative cases:
(i) $U_{eN}=10^{-5}$ and $c_\mathcal{O} = 0$, 
(i) $U_{eN}=0$ and $c_\mathcal{O} = 10^{-5} v^{-2}$, and 
(iii) $U_{eN}=10^{-5}$ and $c_\mathcal{O} = 10^{-5} v^{-2}$.
The value of the operator coefficient ensures 
that the branching ratios are compatible with the errors 
in the measurements presented in the right panel of table~\ref{tab:Kdecays}, most importantly, with that of $K^+ \to e^+ \nu_e$.
The value $U_{eN} = 10^{-5}$ ($|U_{eN}|^2 = 10^{-10}$) is such that it is below the existing constraints.
For HNLs lighter than the kaons, the leading constraints on the mixing parameter come from NA62~\cite{NA62:2020mcv}, PS191~\cite{Bernardi:1987ek}, and T2K~\cite{T2K:2019jwa}, which set upper limits at the level of $|U_{eN}|^2 \sim 5 \times 10^{-10}$.
For even lighter HNLs, below the pion mass, PIENU~\cite{PIENU:2017wbj} has ruled out parameter space corresponding to $|U_{eN}|^2 > 10^{-7} \sim 10^{-8}$.
Moreover, with the chosen value for the mixing, its contribution is comparable to the contributions of most operators for $c_\mathcal{O} \sim 10^{-5} v^{-2}$, \textit{cf.}~eq.~\eqref{eq:cmix}. 
For the scalar operator $\mathcal{O}_{udeN,12}^{S,RR}$, the constraints from NA62 set limits on the operator to $c_\mathcal{O} \sim 10^{-6} v^{-2}$.

\begin{figure}[t]
 %
 \includegraphics[width=0.49\textwidth]{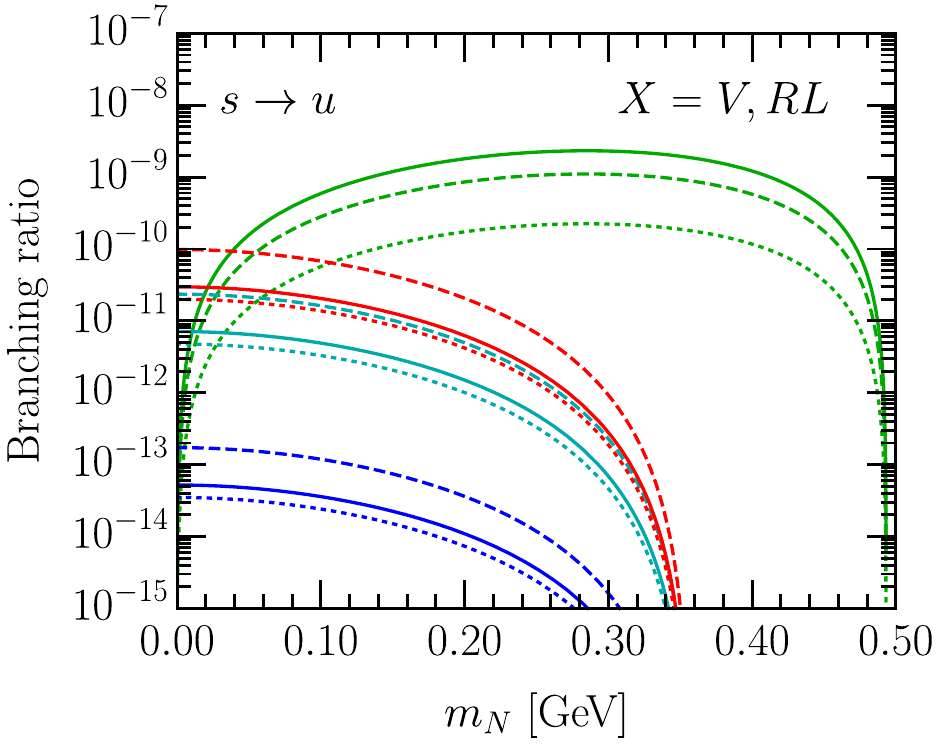}
 \includegraphics[width=0.49\textwidth]{legend_KeN.pdf}\\
 \includegraphics[width=0.49\textwidth]{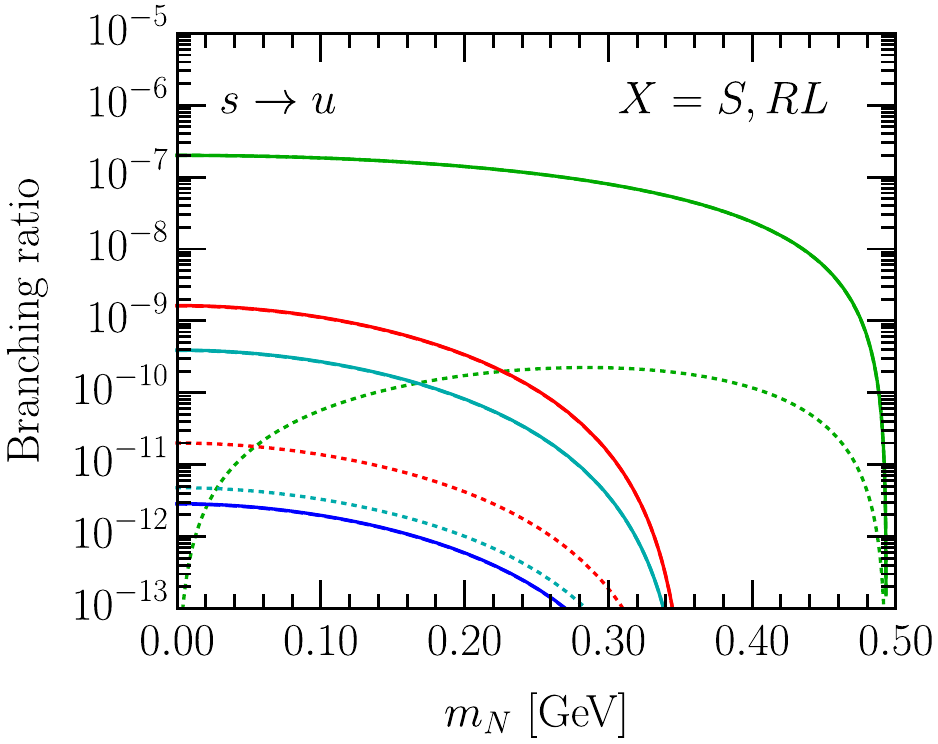} 
 \includegraphics[width=0.49\textwidth]{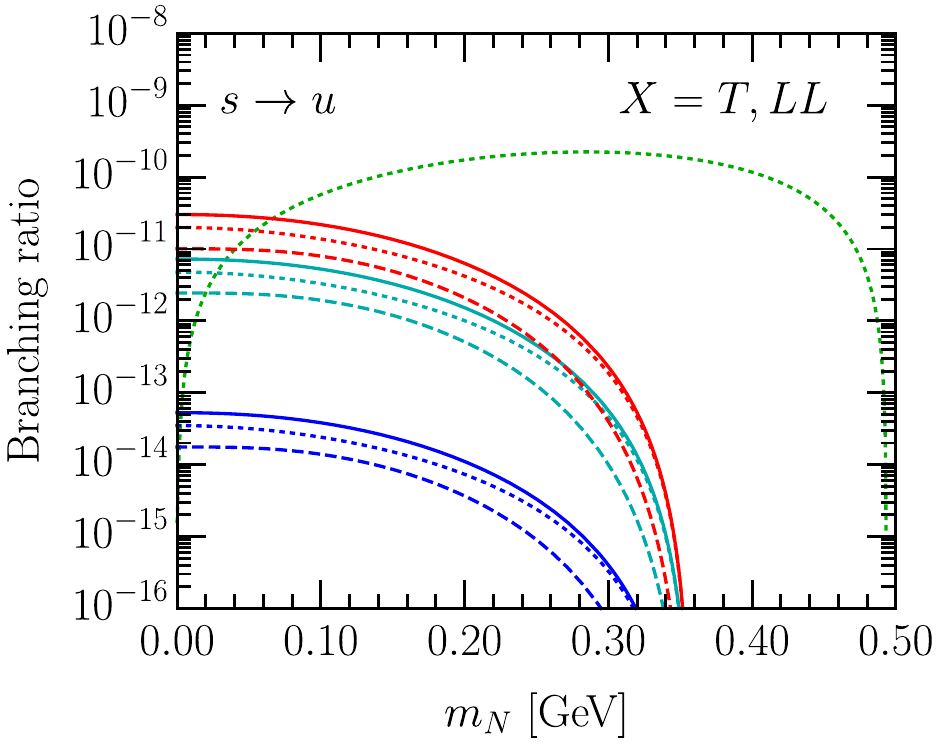}
 \caption{Branching ratios of kaon decays triggered by 
 the LNV single-$N_R$ operators with electron, 
 as well as by active-heavy mixing $U_{eN}$.}
 \label{fig:BRseNLNV}
\end{figure}

In figure~\ref{fig:BRseNLNV}, we show the branching ratios 
for the same processes as in figure~\ref{fig:BRseNLNC}, 
but for the LNV operators
$\mathcal{O}_{udeN,12}^{V,RL}$, 
$\mathcal{O}_{udeN,12}^{S,RL}$, and 
$\mathcal{O}_{udeN,12}^{T,LL}$. 
In the absence of mixing, the results are identical to those for the
corresponding LNC operators (switched on one at a time). 
In the presence of mixing, there is an interference between 
the effective operator generated by new physics and 
the four-fermion interaction (see eq.~\eqref{eq:Lmix}) arising from integrating out the $W$ boson. Its effect is more pronounced for the vector-type operators, and it is stronger for the LNV operator, as can be understood from eq.~\eqref{eq:GammaKeN}.

\section{Experiments and simulation}\label{sec:experiments}

A whole list of far detectors have been proposed for operation in the
vicinity of various interaction points \at{(IPs)} of the LHC, either during Run 3
or the HL-LHC phase; some of them have even been approved and are
running, including FASER and MoEDAL-MAPP1.  Since they are all
sensitive to signatures of tracks stemming from displaced decays of
LLPs taking place inside their fiducial volumes, we will perform the
numerical analysis taking into account these proposals
comprehensively.

We classify these detectors according to their associated IP at the LHC.
For the ATLAS IP, ANUBIS~\cite{Bauer:2019vqk}, FASER~\cite{Feng:2017uoz,FASER:2018eoc}, and FASER2~\cite{Feng:2022inv} are relevant.
ANUBIS\footnote{Very recently, a new design of ANUBIS has been discussed~\cite{ANUBIS_talk_slides,Satterthwaite:2839063}. Instead of being placed inside one of the service shafts, ANUBIS is now considered to be installed at the ATLAS cavern ceiling or shaft bottom.  Given the changing status of the proposal, we stick to the original design for sensitivity study as a good reference for interested readers.} is a detector proposed to be installed inside one of the service shaft above the ATLAS IP.
It has a cylindrical shape of 56~m height and 18~m diameter.
Being close to the ATLAS IP, it is expected to suffer from certain background sources such as neutral kaons.
Nevertheless, exclusion bounds at 95\% confidence level (C.L.) are usually shown with 3 signal events, assuming zero background, and we will follow the practice in this paper.
FASER is currently collecting data at the LHC; first results can be found in Refs.~\cite{FASER:2023zcr,Petersen:2023hgm}.
It is a small cylindrical detector installed in the very forward position along the beam direction, with a distance of 480 m from the IP.
Further, an upgraded program of FASER, FASER2, has been suggested to be built at the site of the proposed Forward Physics Facility~\cite{Feng:2022inv}.
It is larger than FASER and has hence better acceptances to LLPs.
ANUBIS and FASER2 should collect in total 3 ab$^{-1}$ integrated luminosity data, while FASER will have order of 150 fb$^{-1}$ integrated luminosity.

Near the CMS IP, MATHUSLA~\cite{Curtin:2018mvb,Chou:2016lxi,MATHUSLA:2020uve} and FACET~\cite{Cerci:2021nlb} have been proposed.
MATHUSLA would be an experiment on the ground, with about 100 m distance from the CMS IP.
It would have a huge effective volume of $100~\rm{m}\times 100~\rm{m}\times 25~\rm{m}$.
FACET is suggested to be a sub-system of the CMS experiment; with a cylindrical shape, it has a distance of 101~m from the CMS IP, enclosing the beam pipe.
Moreover, with a radius of 0.5~m and a length of 18~m, FACET is relatively large compared to FASER.
Both MATHUSLA and FACET would be running during the HL-LHC and will hence have an integrated luminosity of 3 ab$^{-1}$.

Finally, for the LHCb IP, we have CODEX-b~\cite{Gligorov:2017nwh}, and MoEDAL-MAPP1 and MoEDAL-MAPP2~\cite{Pinfold:2019nqj,Pinfold:2019zwp}.
CODEX-b has been proposed as a cubic box with a dimension of 10~m $\times$ 10~m $\times$ 10~m covering the pseudorapidity range $[0.2,0.6]$, roughly 10~m away from IP8.
MoEDAL-MAPP1 is a small detector of about 140 m$^3$ in a gallery of negative pseudorapidity with respect to the LHCb IP, under operation during Run 3, and MoEDAL-MAPP2 is an enlarged version of MAPP1, to be running during the HL-LHC period. 
The integrated luminosity associated with the LHCb IP is lower compared to that of ATLAS or CMS.
CODEX-b and MoEDAL-MAPP2 will have an integrated luminosity of 300 fb$^{-1}$ while MoEDAL-MAPP1 has only 30 fb$^{-1}$.

For a more detailed summary of these proposals, we refer the reader to \textit{e.g.}~Refs.~\cite{Lee:2018pag,Alimena:2019zri,DeVries:2020jbs}, as well as to the respective references proposing the detectors.

In addition, at DUNE, a proton beam of energy 120 GeV hits a fixed target with $1.1\times 10^{21}$ POTs (protons on target) expected per year.
The produced mesons decay either promptly or with a macroscopic distance into the LLPs (which are the HNLs in this study) inside a decay pipe which is 26 m downstream from the IP and has a length of 194 m and a radius of 2 m.
The produced LLPs should then travel a long distance before decaying inside the DUNE-ND which is further downstream with a distance of 574 m from the fixed target, and has a length of 6.4 m and a width of 3.5 m.
We take an operation duration of 10 years as benchmark in this study, expecting thus in total $1.1\times 10^{22}$ POTs.

In order to obtain the kinematics of the HNLs produced from kaons, we make use of the tool Pythia8 with the module \textit{SoftQCD:all}.
The kaons are generated in $pp$ collisions with a center-of-mass-energy 14 TeV, and are set to decay exclusively in the signal-event channels.
Pythia8 provides the boost factor and the polar angle of each simulated HNL.
We note that since kaons are themselves long-lived, we let Pythia8 decide the decay positions of the kaons and thus take into account the production position of the HNL ($=$ the decay position of the kaon) as well as the polar angle and boost factor of the HNL, in the computation of the HNLs' average decay probability inside the far detectors.
The total signal-event rates at each detector can be computed with the following formula
\begin{eqnarray}
N_S =   \sum_{K's} N_K \cdot n\cdot \text{BR}\left(K\to n \, N+ \text{anything}\right) \cdot \epsilon \cdot \text{BR}\left(N\to \text{visible}\right)\,,	
\end{eqnarray}
where the summation goes over different kaons, $N_K$ is the total number of kaons, $n=1,2$ is the number of HNLs produced in the considered kaon decays,  BR$(N\to\text{visible})$ is the visible decay branching ratio of the HNL, and $\epsilon$ is the average decay probability in a far detector.
For the LHC experiments, we estimate the number of kaons $N_K$, with the help of the tool EPOS LHC~\cite{Pierog:2013ria} provided in the CRMC simulation package~\cite{CRMC}, to be $N_{K^\pm}=2.38\times 10^{18}$, $N_{K_S}=1.31\times 10^{18}$, and $N_{K_L}=1.30\times 10^{18}$, 
over the whole $4\pi$ solid angle.
For the DUNE experiment, we follow Refs.~\cite{Krasnov:2019kdc,Gunther:2023vmz} and conclude that the total numbers of kaons produced at DUNE for 10 years are $N_{K^\pm} = N_{K_S} = N_{K_L} =5.76\times 10^{21}$ over the whole $4\pi$ solid-angle range.
The computation procedure of $\epsilon$ is based on the exact formulae given in Refs.~\cite{Dercks:2018eua,Dercks:2018wum,DeVries:2020jbs,Hirsch:2020klk,Beltran:2022ast} and the further development presented in Ref.~\cite{Gunther:2023vmz} for LLPs from kaons.
As the kaons, especially $K^\pm$ and $K_L$, travel macroscopic distances, the infrastructure surrounding the IPs may affect the kinematics of the kaons.
To simplify the analysis, we neglect the influence of any magnetic fields present at the LHC IPs or the magnetic horns at DUNE.
Additionally, we introduce cut-offs for the production positions of the HNL at the LHC that are included in the signal-event rates $N_S$, resulting in conservative estimates.
Here, we provide a brief summary of the cut-offs we employ in our simulation; for more detail, see Ref.~\cite{Gunther:2023vmz}.
For example, a lead shield covering the total fiducial volume in order to veto neutral background events is placed approximately \SI{5}{\meter} in front of CODEX-b.
Hence, we require the kaons to decay before reaching the shield.
For detectors in the far forward region (FASER, FASER2, and FACET), we employ the beamline geometry of the ATLAS and CMS IP, which involve absorbers for charged and neutral particles in order to protect beamline infrastructure.
The hadron calorimetry of ATLAS and CMS restricts the decay region of the kaons for ANUBIS and MATHUSLA, respectively.
Lastly, for MoEDAL-MAPP1 and MAPP2, the natural rock between the IP and the detectors is the limiting factor.
Hence, the kaons are required to decay within the \SI{3.8}{\meter} wide beam cavern.
We note that such cut-offs are not needed for the DUNE-ND, as a decay pipe for the long-lived mesons to decay in should be instrumented.

Finally, we discuss the procedure we apply for recasting the bounds on the HNLs in the minimal scenario, obtained in some past searches, into those on the HNLs in the EFT scenarios considered here.
In general, we follow the approaches laid out in Refs.~\cite{Beltran:2023nli,Fernandez-Martinez:2023phj,Dreiner:2023gir} (see also Ref.~\cite{Barouki:2022bkt}).
We consider three searches at NA62~\cite{NA62:2020mcv}, PS191~\cite{Bernardi:1987ek}, and T2K~\cite{T2K:2019jwa}.
Since these searches all require a prompt charged lepton, they are applicable only to the single-$N_R$ scenarios.
We first consider the NA62 search, which looked for HNL production in $K^+$ decays to positrons and missing energy, assuming the proper lifetime of $N$ is larger than 50 ns.
The search obtained bounds on BR$(K^+ \to e^+ N)$ and hence those on the active-sterile neutrino mixing, as functions of the sterile neutrino mass.
We simply convert BR$(K^+ \to e^+ N)$ to the production Wilson coefficient of the single-$N_R$ scenario in question, for each mass value, and obtain the corresponding recast bounds.
Both the PS191 and T2K searches are for both a prompt charged lepton and a displaced vertex at the detached detector.
For PS191, we extract the sensitivity curve presented in the plane $|U_{eN}|^2$ vs.~$m_N$ for the signal process $K^+ \to e^+ N$, $N \to e^- \pi^+$ and its charge-conjugate channel, and for the T2K near detector ND280, the signal process $K^\pm \to e^\pm N$, $N \to e^\pm \pi^\mp$ is considered.
By rescaling the production and visible decay rates, 
we obtain the recast bounds in the EFT parameter space.

\section{Results}
\label{sec:results}

For presenting the numerical results, we choose several benchmarks 
characterized by the different couplings responsible 
for the HNL production and decay. 
We summarize them in table~\ref{tab:benchmarks}.
\begin{table}[t]
\renewcommand{\arraystretch}{1.2}
\centering
 \begin{tabular}[t]{|l|l|l|}
 \hline
 Benchmark & Production & Decay \\
 \hline
 B1.1 & $c_{dN,21}^{V,RR} \in \mathbb{R}$ & $U_{eN}$ \\
 B1.2 & $c_{dN,21}^{V,RR} \in i\, \mathbb{R}$  & $U_{eN}$ \\
 \hline
 B2.1 & $c_{dN,21}^{S,RR} \in \mathbb{R}$ & $U_{eN}$ \\
 B2.2 & $c_{dN,21}^{S,RR} \in i\,\mathbb{R}$ & $U_{eN}$ \\
 \hline
 \end{tabular}
 \hspace{0.5cm}
 \begin{tabular}[t]{|l|l|l|}
 \hline
 Benchmark & Production & Decay \\
 \hline
 B3 & $c_{udeN,12}^{V,RR}$ & $c_{udeN,11}^{V,RR}$ \\
 B4 & $c_{udeN,12}^{S,RR}$ & $c_{udeN,11}^{S,RR}$ \\
 \hline
 B5 & $c_{udeN,12}^{V,RR}$ and $U_{eN}$ & $U_{eN}$ \\
 B6 & $c_{udeN,12}^{S,RR}$ and $U_{eN}$ & $U_{eN}$ \\
 B7 & $c_{udeN,12}^{V,RL}$ and $U_{eN}$ & $U_{eN}$ \\
 B8 & $c_{udeN,12}^{S,RL}$ and $U_{eN}$ & $U_{eN}$ \\
 \hline
 \end{tabular}
\caption{Benchmarks for the scenarios with pair-$N_R$ (left) 
and single-$N_R$ (right) operators.}
\label{tab:benchmarks}
\end{table}
In total, we consider four scenarios with pair-$N_R$ operators 
(benchmarks B1 and B2, each having two sub-cases) 
and six scenarios with single-$N_R$ operators (benchmarks B3--B8).

\begin{figure}[t]
 \centering
 \includegraphics[width=0.49\textwidth]{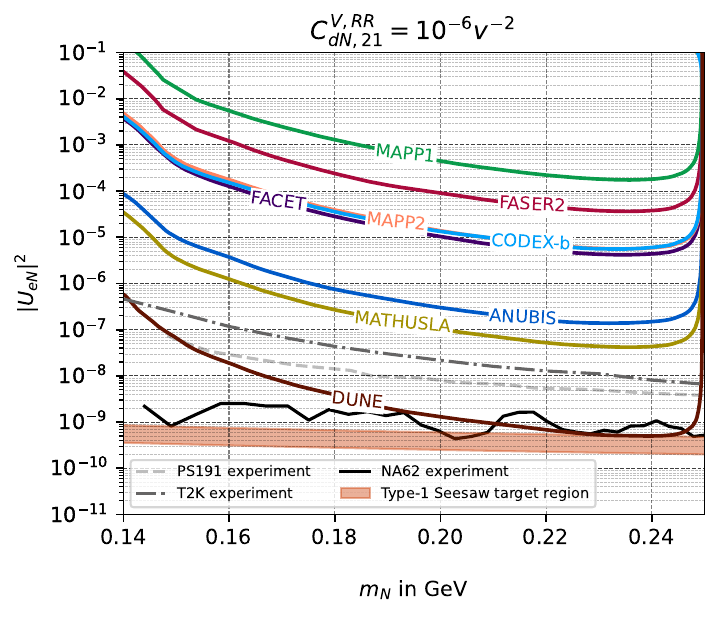}
 \hfill
 \includegraphics[width=0.49\textwidth]{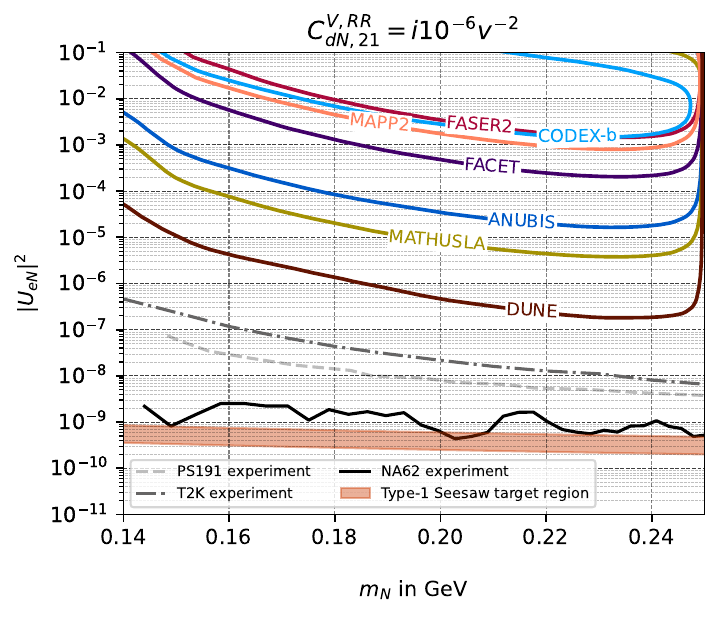}\\
 \includegraphics[width=0.49\textwidth]{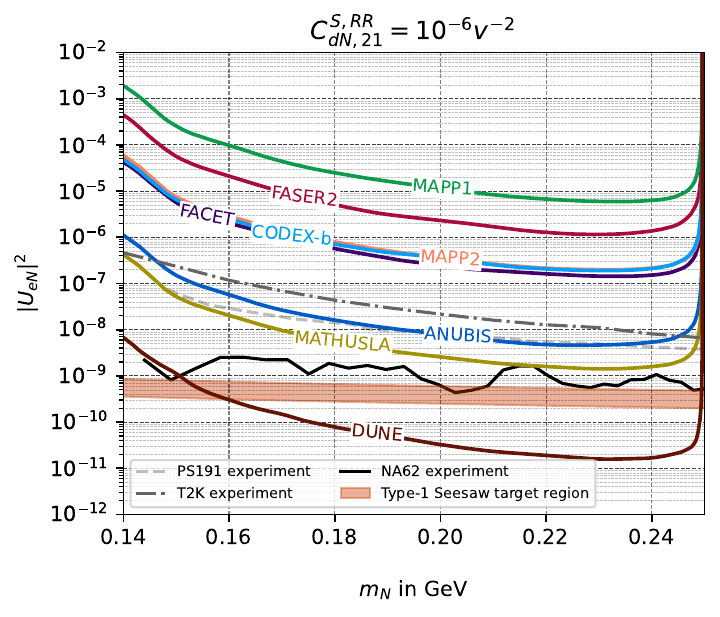}
 \hfill
 \includegraphics[width=0.49\textwidth]{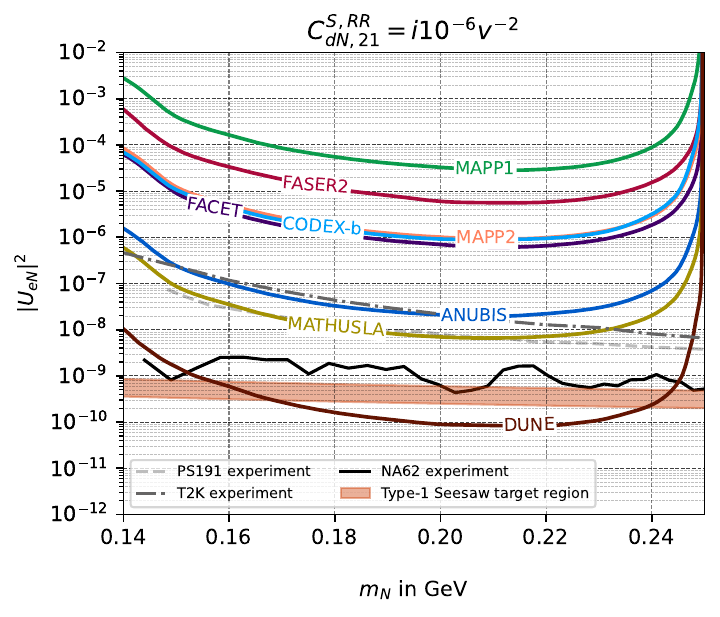}\\
 \caption{Projected exclusion limits in the plane $|U_{eN}|^2$ vs. $m_N$ 
 for pair-$N_R$ operator benchmarks B1.1 and B1.2 (top), 
 and B2.1 and B2.2 (bottom). 
 The absolute value of the corresponding WC has been fixed to $10^{-6} v^{-2}$.
 The current bounds from NA62, T2K, and PS191, as well as 
 the type-I seesaw target region where $m_\nu=0.05$--$0.12$~eV, are also shown.}
 \label{fig:B1B2fixedWC}
\end{figure}

For benchmark B1 (B2), HNL production is governed 
by the LNC (LNV) pair-$N_R$ operator $\mathcal{O}_{dN,21}^{V,RR}$ 
($\mathcal{O}_{dN,21}^{S,RR}$), while HNL decay proceeds via active-heavy neutrino mixing $U_{eN}$. For each of these two operators, 
we consider two cases: 
(i) real WC and (ii) purely imaginary WC, \textit{cf.}~figure~\ref{fig:BRsNN} and the related discussion. 
First, in figure~\ref{fig:B1B2fixedWC}, we fix the absolute value of the WC 
to $10^{-6} v^{-2}$ (to respect the constraint coming from $K^+ \to \pi^+ \nu \overline{\nu}$) and display the projected exclusion limits 
in the plane $|U_{eN}|^2$ vs. $m_N$ for three signal events.
The curves correspond to 95\% C.L. limits (under the assumption of zero background). 
For the LNC operator, the sensitivities are approximately 
two orders of magnitude weaker in the case of purely imaginary WC 
(B1.2, top-right plot) than in the case of real WC of the same size 
(B1.1, top-left plot). This can be understood from figure~\ref{fig:BRsNN}, 
which shows that for the same size of the WC, 
BR$(K_S \to NN)$ is about two orders of magnitude smaller 
than BR$(K_L \to NN)$, whereas the production numbers of $K_S$ and $K_L$ 
are nearly the same. The difference in the sensitivities 
is much milder for the LNV operator, 
for which the branching ratios of the relevant kaon decays differ 
only slightly between the cases of real and purely imaginary WC.

\begin{figure}[t]
 \centering
 \includegraphics[width=0.49\textwidth]{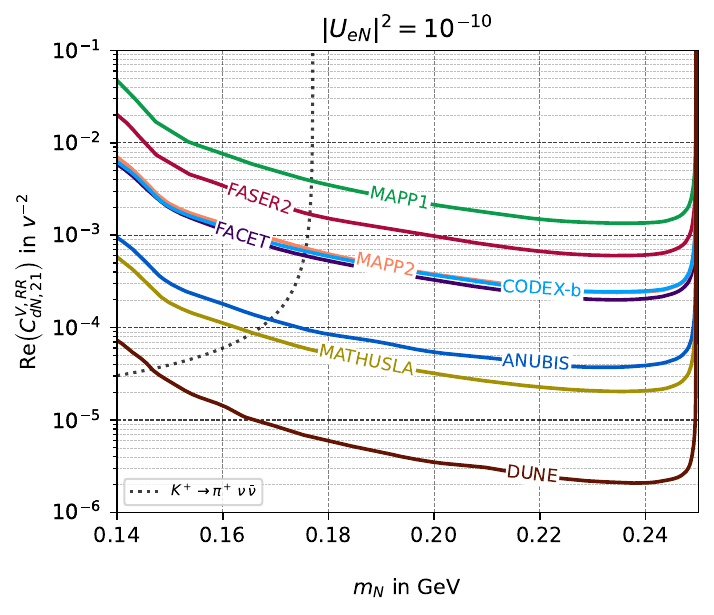}
 \hfill
 \includegraphics[width=0.49\textwidth]{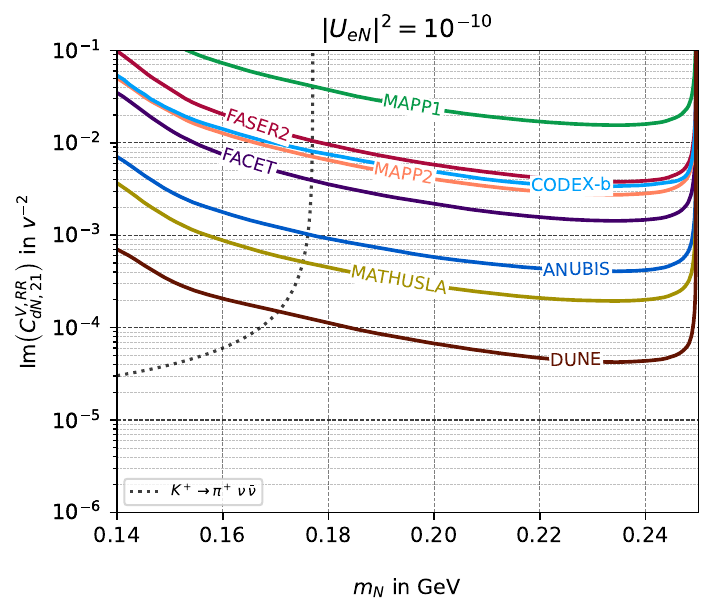}\\
 \includegraphics[width=0.49\textwidth]{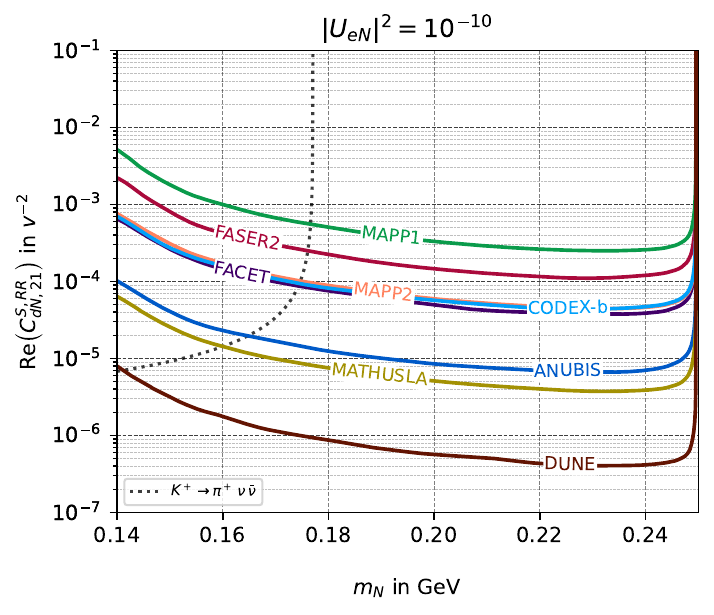}
 \hfill
 \includegraphics[width=0.49\textwidth]{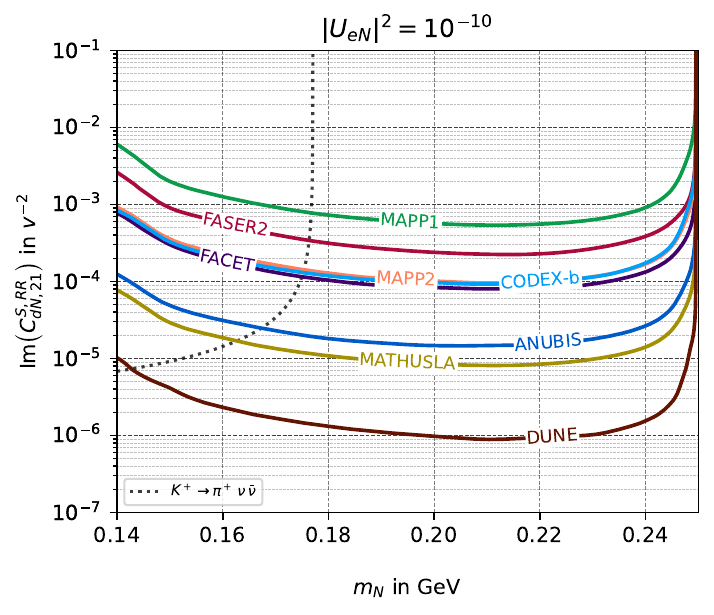}\\
 \caption{Exclusion limits in the plane WC vs.~$m_N$ 
 for pair-$N_R$ operator benchmarks B1.1 and B1.2 (top), 
 and B2.1 and B2.2 (bottom). 
 The active-heavy neutrino mixing parameter has been fixed as 
 $|U_{eN}|^2 = 10^{-10}$. 
 The dotted line represents the constraint
 originating from the measured branching ratio of $K^+ \to \pi^+ \nu \overline{\nu}$.}
 \label{fig:B1B2fixedU}
\end{figure}

Among the considered LLP detectors, the best limit comes from MATHUSLA, 
which can probe $|U_{eN}|^2$ down to $4\times10^{-8}$ ($1.4\times10^{-9}$) 
at $m_N \approx 0.23$~GeV for B1.1 (B2.1). 
It is followed by ANUBIS, which has a factor of a few weaker sensitivity. 
MAPP2, FACET and CODEX-b provide very similar exclusion limits, 
which are approximately two orders of magnitude weaker than the expected limits from MATHUSLA. 
Finally, FASER2 and MAPP1 have the weakest depicted sensitivities.\footnote{For FASER, 
there is no isocurve shown in figure~\ref{fig:B1B2fixedWC} and figure~\ref{fig:B1B2fixedU}. Because of the small geometric acceptance and comparatively lower integrated luminosity, the simulated number of signal events is less than three in the shown parameter region.}
However, all the LHC far detectors are incomparable to the DUNE-ND which can probe $|U_{eN}|^2$ down to the levels stronger than MATHUSLA by up to two orders of magnitude in these scenarios.
This superior performance is mainly due to the much larger production rates of the kaons at DUNE.
We also show the existing limits on active-heavy neutrino mixing 
obtained by the NA62~\cite{NA62:2020mcv}, T2K~\cite{T2K:2019jwa}, and PS191~\cite{Bernardi:1987ek} experiments. 
Though derived under the hypothesis of the minimal mixing case, 
these limits apply to the considered EFT scenarios as well. 
As can be seen, NA62 outperforms all the LHC far detectors 
and already touches the (naive) type-I seesaw band, 
where the values of $m_N$ and $U_{eN}$ yield the light neutrino mass 
$m_\nu = 0.05$--$0.12$~eV.
We find that the DUNE-ND can still be sensitive to parameter space beyond the current bounds, especially in the scenarios of the benchmark B2, where it covers the type-I seesaw band.

In figure~\ref{fig:B1B2fixedU}, we fix $|U_{eN}|^2 = 10^{-10}$ 
and show the exclusion limits in the plane WC vs.~$m_N$.
For benchmark B1.1 (B1.2), MATHUSLA will be able to probe 
the WC as small as $2\times10^{-5} v^{-2}$ ($ 2 \times 10^{-4} v^{-2}$) for $m_N$ in the range 0.22--0.24~GeV. 
These numbers translate to the new physics scale $\Lambda$ of 55 (17)~TeV.%
\footnote{We associate the new physics scale $\Lambda$ with 
the effective operators in the $N_R$SMEFT, assuming 
their WCs are $\Lambda^{4-d}$, 
with $d$ denoting the operator mass dimension.
The LNC four-fermion operators in the $N_R$LEFT arise from 
$d=6$ four-fermion operators in the $N_R$SMEFT, 
whereas the LNV four-fermion operators in the $N_R$LEFT 
originate from $d=7$ operators in the $N_R$SMEFT.
For details of the matching between the two effective theories, 
see \textit{e.g.}~Refs.~\cite{Beltran:2022ast,DeVries:2020jbs,Dekens:2020ttz}.} 
For DUNE-ND, the sensitivity reach can be further down to $2\times 10^{-6} v^{-2}$ $(4\times 10^{-5} v^{-2})$, corresponding to 174 (39) TeV.
In the case of the LNV operator, we find that for benchmark B2.1 (B2.2), 
MATHUSLA can exclude the effective couplings down to $4\times10^{-6} v^{-2}$ for $m_N \approx 0.23$~GeV 
(8$\times10^{-6}v^{-2}$ for $m_N \approx 0.21$~GeV). However, being an LNV operator, it originates at
 dimension 7 in the $N_R$SMEFT, and hence, these exclusion limits 
translate to $\Lambda$ of around 11 (9)~TeV.
DUNE-ND performs better than the LHC far detectors again, reaching $4\times 10^{-7} v^{-2}$ ($8\times 10^{-7} v^{-2}$) in the benchmark B2.1 (B2.2), corresponding to the new-physics scales of around 24 TeV (19 TeV) at $m_N=0.23$ GeV (0.21 GeV).
We also show the constraint originating from the measurement of the branching ratio of $K^+ \to \pi^+ \nu \overline{\nu}$. 
It is complementary to the projected limits for $m_N \lesssim 0.16$--0.17~GeV. 
We recall that the NA62, PS191, and T2K exclusion limits on active-heavy mixing cannot be reinterpreted into the limits on the WCs of the pair-$N_R$ operators, since these interactions do not lead to a prompt charged lepton.

\begin{figure}[t]
 \centering
 \includegraphics[width=0.49\textwidth]{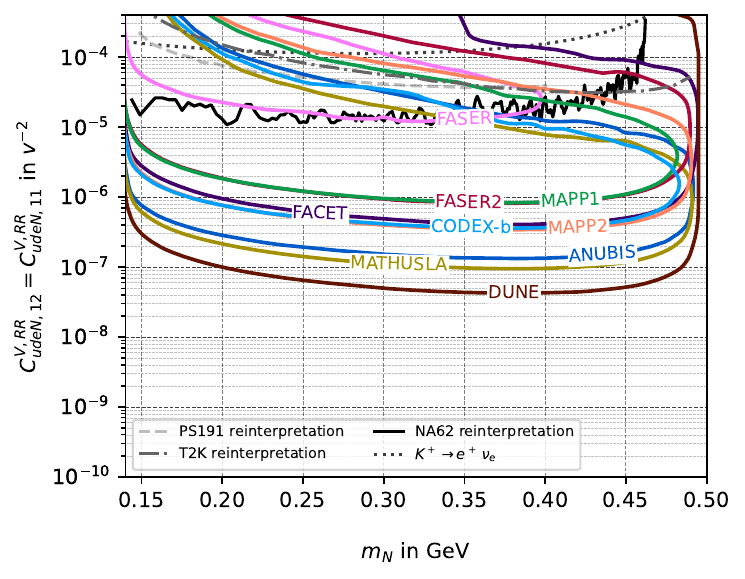}
 \hfill
 \includegraphics[width=0.49\textwidth]{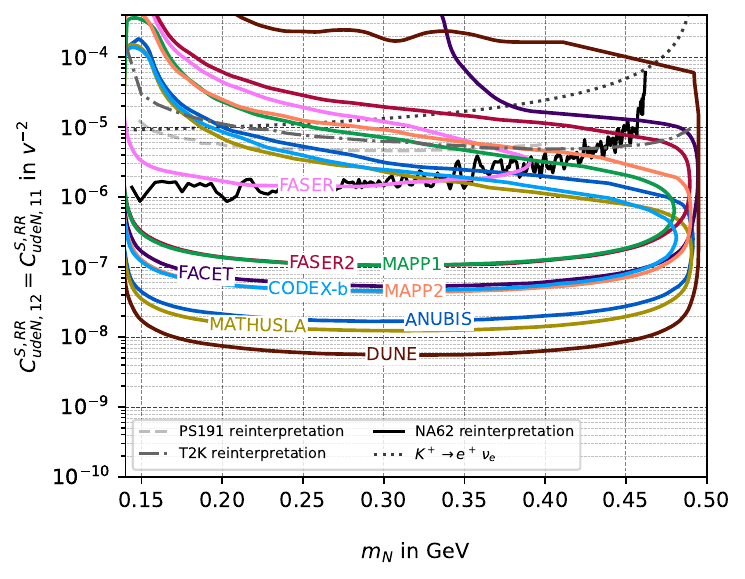}\\
 \caption{Exclusion limits in the plane WC vs.~$m_N$ for single-$N_R$ operator benchmarks B3 (left) 
 and B4 (right). The production and decay couplings have been assumed to be equal. 
 The recast bounds from NA62, T2K, and PS191, as well as 
 the constraint originating from the measured branching ratio of 
 $K^+ \to e^+ \nu_e$, are also shown.}
 \label{fig:B3B4}
\end{figure}

Next, we consider the single-$N_R$ operator benchmarks 
summarized in the right part of table~\ref{tab:benchmarks}. 
We start with benchmarks B3 and B4, for which we assume 
that both HNL production and decay proceed via the same effective operator, 
but carrying different quark flavor indices. 
The indices 12 lead to the HNL production in kaon decays, 
while the indices 11 realize the decay $N \to e^\mp \pi^\pm$. 
For these benchmarks, we also assume that 
there is no active-heavy neutrino mixing. 
We consider two single-$N_R$ operators $\mathcal{O}_{udeN}^{V,RR}$ 
and $\mathcal{O}_{udeN}^{S,RR}$, both conserving the lepton number. 
(The results for the corresponding LNV operators are the same 
under the assumption of zero active-heavy neutrino mixing.)
For graphical presentation, we choose to set the production and decay 
couplings equal, \textit{i.e.}~$c_{udeN,12}^{V,RR} = c_{udeN,11}^{V,RR}$ for B3, and $c_{udeN,12}^{S,RR} = c_{udeN,11}^{S,RR}$ for B4, 
and show the derived exclusion limits in the plane WC vs.~$m_N$,
see figure~\ref{fig:B3B4}.
In addition to the sensitivities of the proposed LLP detectors at the LHC and the DUNE-ND, 
we show the recast bounds from NA62~\cite{NA62:2020mcv}, T2K~\cite{T2K:2019jwa}, and PS191~\cite{Bernardi:1987ek}, 
obtained according to the procedure explained at the end of Sec.~\ref{sec:experiments}. 
Except for FASER, of which the sensitivity is comparable to that of NA62, all the far detectors at the LHC and the DUNE-ND will have better reach to these scenarios than NA62, which excludes WC values larger than approximately (1--2)$\times10^{-5}v^{-2}$ ((9--20)$\times10^{-7}v^{-2}$) for benchmark B3 (B4).
For the vector-type operator and $m_N \approx 0.35$--$0.40$~GeV, DUNE-ND and MATHUSLA will probe the effective couplings as small as $4\times 10^{-8} v^{-2}$ and $9.5\times10^{-8} v^{-2}$, respectively, and FASER down to $1.4\times10^{-5} v^{-2}$, 
with the sensitivities of the other experiments lying between these extremes.
Translating these numbers to the new-physics scale $\Lambda$, we find that DUNE (MATHUSLA, FASER) will be sensitive to 
$\Lambda$ as high as 1230 (798, 65) TeV. 
For the scalar-type operator, the reach in the $N_R$LEFT WC is around one order of magnitude better, owing to the 
larger branching ratio of $K^\pm \to e^\pm N$ in this case, see figure~\ref{fig:BRseNLNC}. 
The new-physics scales, which could be probed by DUNE (MATHUSLA, FASER) for $m_N \approx0.25$--0.35~GeV, are in excess 
of 3000 (2000, 200)~TeV. 
The projected limits from DUNE and MATHUSLA are approximately more than ten times more stringent than the limits derived on the effective interactions 
in Ref.~\cite{Zhou:2021lnl} from the LNV decays $K^{\mp} \to \pi^\pm \ell^\mp \ell^\mp$ mediated by $N$.

\begin{figure}[t]
 \centering
 \includegraphics[width=0.49\textwidth]{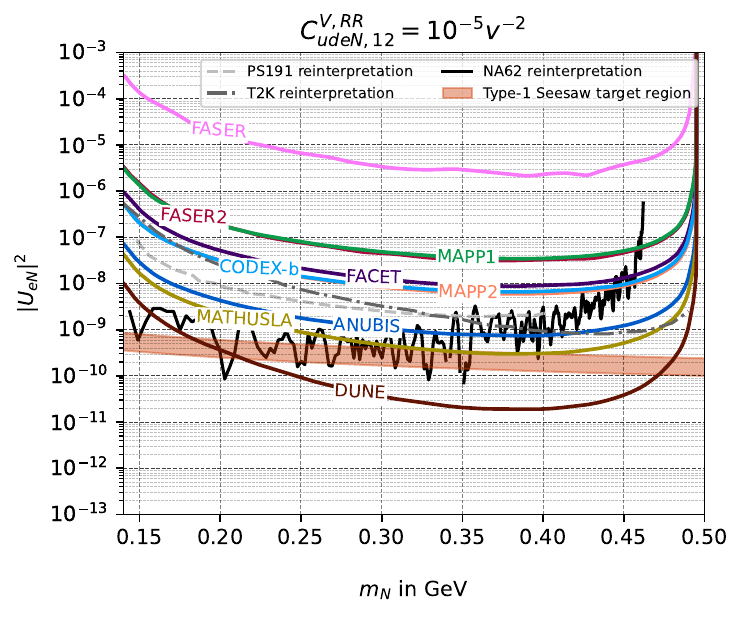}
 \hfill
 \includegraphics[width=0.49\textwidth]{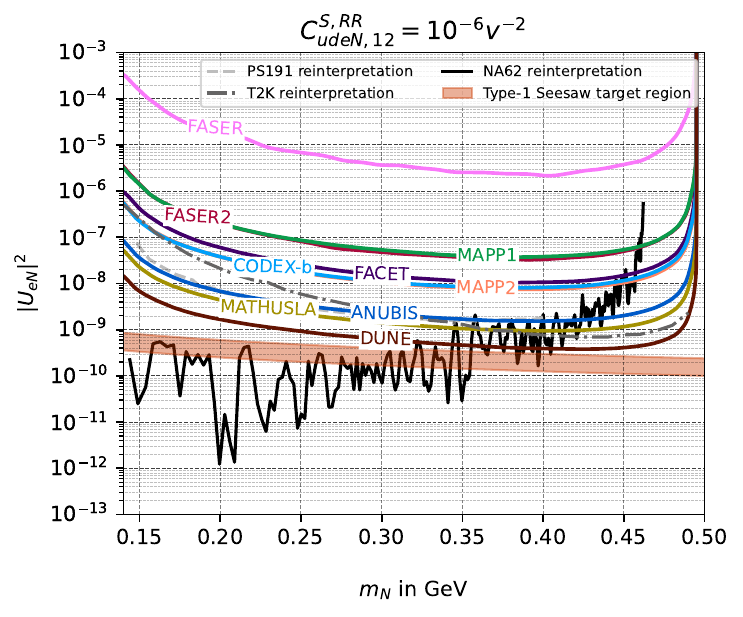}\\
 \includegraphics[width=0.49\textwidth]{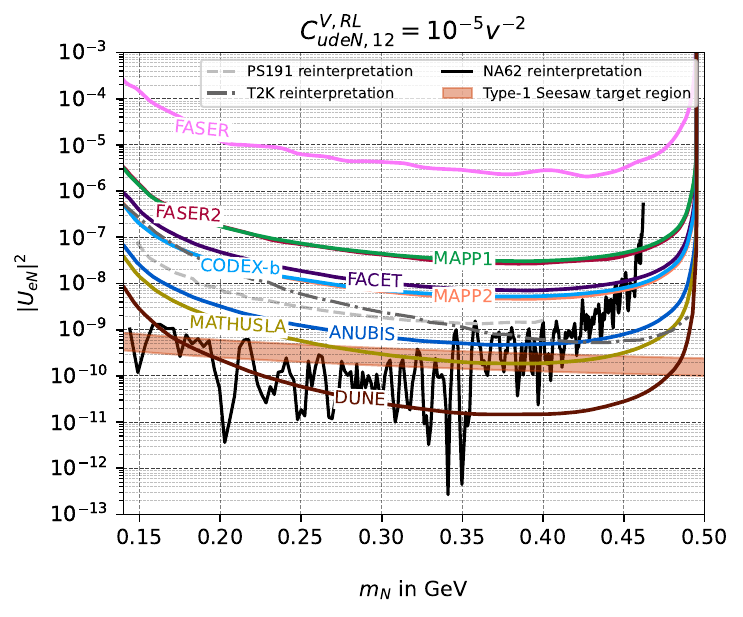}
 \hfill
 \includegraphics[width=0.49\textwidth]{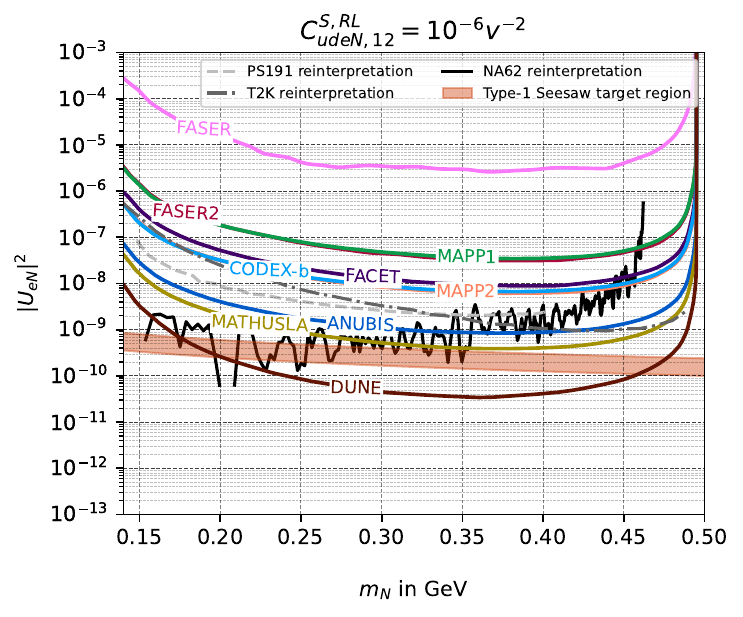}\\
 \caption{Projected exclusion limits in the plane $|U_{eN}|^2$ vs. $m_N$ 
 for single-$N_R$ operator benchmarks B5 and B6 (top), 
 and B7 and B8 (bottom). 
 The absolute value of the corresponding WC has been fixed to $10^{-5} v^{-2}$ 
 and $10^{-6} v^{-2}$ for vector and scalar operators, respectively. 
 The recast bounds from NA62, T2K, and PS191 are also shown.}
 \label{fig:B5B8fixedWC}
\end{figure}

Finally, we turn to benchmarks B5--B8, where both an effective operator and active-heavy neutrino mixing contribute to HNL production, while the HNL decay proceeds via mixing only, see table~\ref{tab:benchmarks}. 
Here, we consider both LNC vector (B5) and scalar (B6) operators, and LNV vector (B7) and scalar (B8) operators, since the interference between the effective interaction and the mixing term is slightly different in the LNC and LNV cases, \textit{cf.}~figures~\ref{fig:BRseNLNC} and \ref{fig:BRseNLNV}.
In figure~\ref{fig:B5B8fixedWC}, we fix the corresponding WC to either $10^{-5} v^{-2}$ for vector operators (to satisfy the constraint coming from $K^+ \to e^+ \nu_e$), or $10^{-6} v^{-2}$ for scalar operators (to satisfy the recast bounds from NA62), and present the sensitivities in the plane $|U_{eN}|^2$ vs.~$m_N$.
For the vector-type operators and $m_N \approx 0.35$--$0.40$~GeV, MATHUSLA could probe $|U_{eN}|^2$ down to $3.0~(2.0) \times 10^{-10}$ for the LNC (LNV) operator, while DUNE can be sensitive to the $|U_{eN}|^2$ values as low as $2\times 10^{-11}$ ($1.5\times 10^{-11}$).
For the scalar-type operators in the same $m_N$-range, the MATHUSLA reach is $|U_{eN}|^2 \approx 9.0~(4.0)\times 10^{-10}$ for the LNC (LNV) operator and the DUNE reach to $|U_{eN}|^2$ is down to approximately $4\times 10^{-10}$ ($3\times 10^{-11}$).

We also display the recast bounds from NA62, T2K, and PS191. 
The NA62 limit is more stringent than the expected limits from the future LLP detectors for $m_N \lesssim 0.35~(0.41)$~GeV in B5 and B8 (B6 and B7). 
For larger $m_N$, MATHUSLA takes over, excluding new parts of the parameter space. 
In particular, for B7, the projected MATHUSLA exclusion limits can probe the seesaw target region for $0.25~\text{GeV} \lesssim m_N \lesssim 0.45$~GeV.
DUNE can, however, probe $|U_{eN}|^2$ values smaller than the current bounds by up to about 2 orders of magnitude, in B5, B7, and B8 benchmarks.
For B6, (only) DUNE can exclude a small part of the parameter space at $m_N\gtrsim 0.4$ GeV.

\begin{figure}[t]
 \centering
 \includegraphics[width=0.49\textwidth]{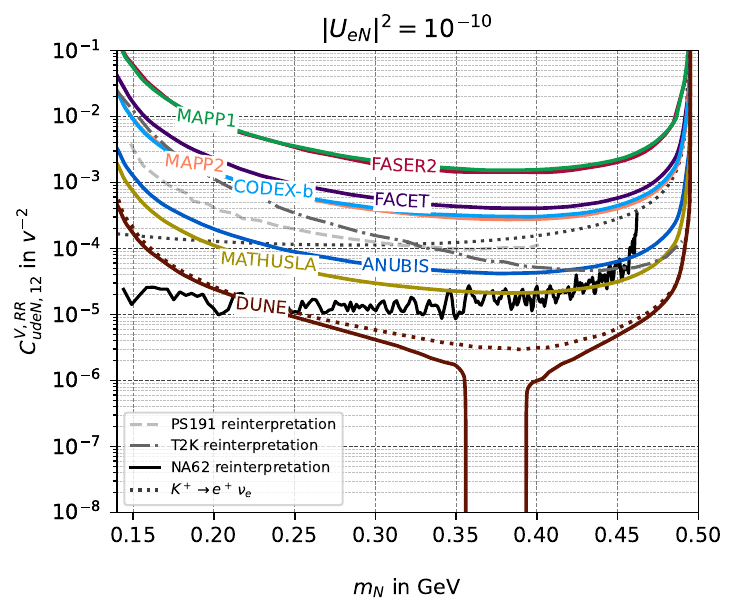}
 \hfill
 \includegraphics[width=0.49\textwidth]{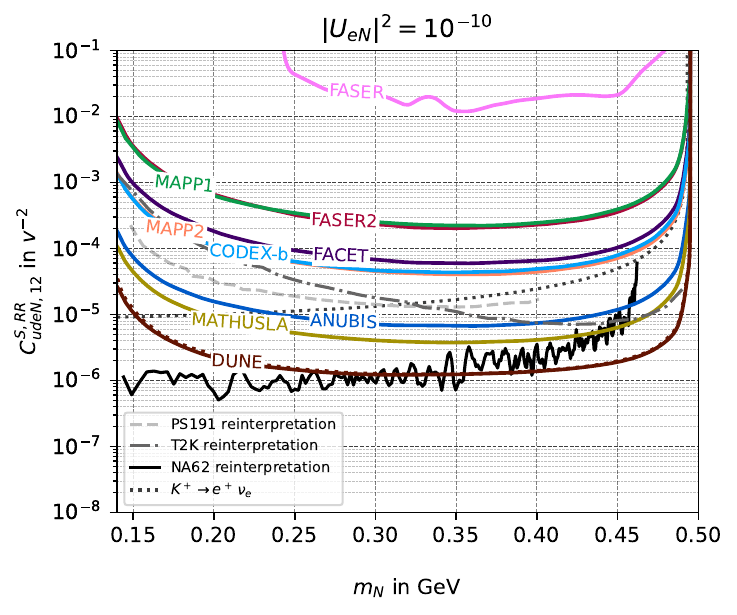}\\
 \includegraphics[width=0.49\textwidth]{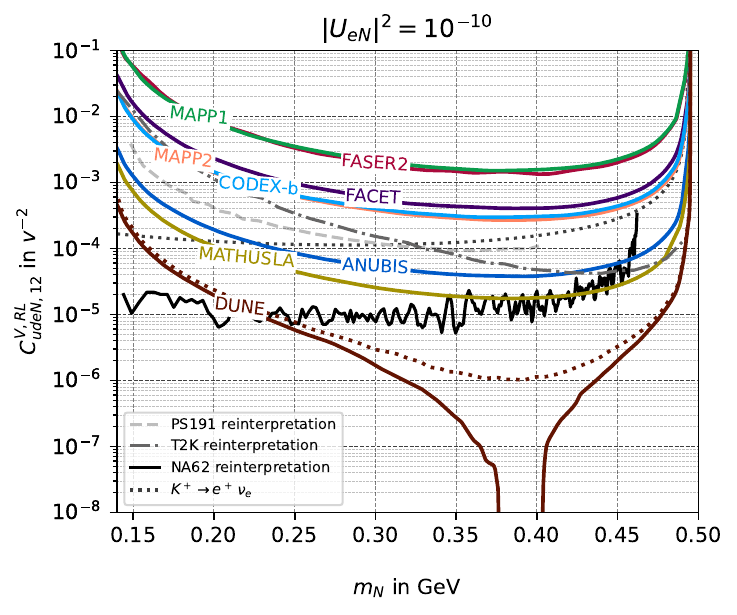}
 \hfill
 \includegraphics[width=0.49\textwidth]{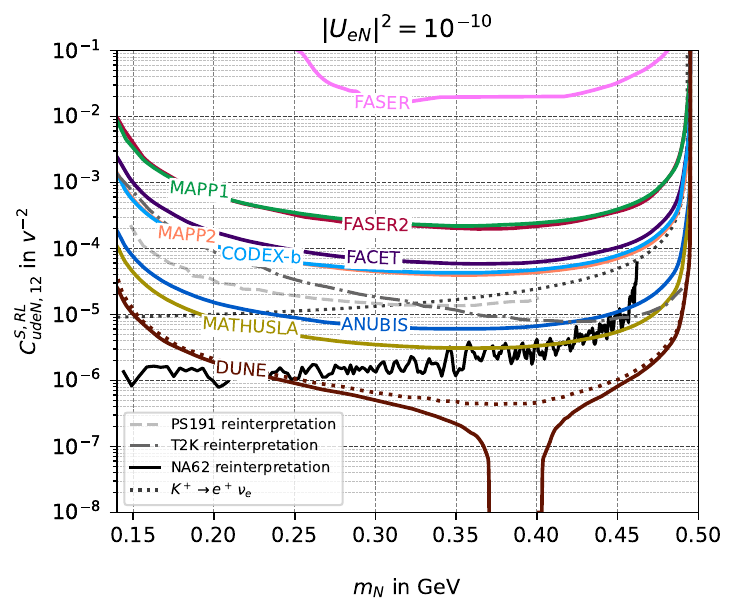}\\
 \caption{Exclusion limits in the plane WC vs. $m_N$ 
 for single-$N_R$ operator benchmarks B5 and B6 (top), 
 and B7 and B8 (bottom). 
 The active-heavy neutrino mixing parameter has been fixed as 
 $|U_{eN}|^2 = 10^{-10}$. 
 The recast bounds from NA62, T2K, and PS191, as well as 
 the constraint originating from the measured branching ratio of 
 $K^+ \to e^+ \nu_e$, are also shown.
For the predictions on the sensitivities of DUNE (brown), we show an extra curve in dotted line style corresponding to $|U_{eN}|^2=8\times 10^{-11}$.
}
 \label{fig:B5B8fixedU}
\end{figure}

In figure~\ref{fig:B5B8fixedU}, we fix $|U_{eN}|^2 = 10^{-10}$ and show the exclusion limits in the plane WC vs.~$m_N$.
We again depict recast bounds from NA62, T2K, and PS191.
The recast NA62 bound covers the ranges which will be accessible to
the future LLP detectors for $m_N \lesssim 0.45$~GeV.  For larger HNL
masses, MATHUSLA will probe an unexplored region of the parameter
space, ruling out the WC $\gtrsim 2 \times 10^{-5} v^{-2}$ ($\gtrsim 3
\times 10^{-6} v^{-2}$) for the vector-type (scalar-type) operators.
On the other hand, the DUNE-ND can exclude new
  parameter space for $m_N\gtrsim 0.25$ (0.35) GeV in benchmarks B5,
  B7, and B8 (B6), showing again the much better constraining power
  than the future LLP far detectors. In the plots
  for the benchmarks B5, B7, and B8, we observe a unique funnel
  feature in the DUNE sensitivity reach for $m_N$ roughly between 0.36
  GeV and 0.4 GeV.  This arises from the fact that in this mass range
  with $|U_{eN}|^2 = 10^{-10}$, even in the absence of the effective
  operators considered, more than three signal events are predicted at
  DUNE (see figure 6 of
  Ref.~\cite{Gunther:2023vmz}),\footnote{We stress that the
      apparent sensitivity at small values of the operator coefficients is feigned. There is no real sensitivity to such small coefficients, since the event number does not depend on
      the operator anymore.}  as long as the interference between the
  minimal mixing and the EFT operators for the HNL production is
  constructive.  Indeed, the funnel feature of the DUNE sensitivities
  do not appear in B6 (the upper right plot), exactly as a result of
  the destructive interference in this case. To
  illustrate this effect better, we choose to show, in all these four
  plots, an additional sensitivity curve for DUNE corresponding to
  $|U_{eN}|^2=8\times 10^{-11}$ which is slightly smaller than the
  default $10^{-10}$ value we have chosen.  We now observe that the
  funnel feature in B5, B7, and B8 has disappeared, and in the plot
  for B6, we find the two DUNE curves for $|U_{eN}|^2 = 10^{-10}$ and
  $8\times 10^{-11}$ almost completely overlap as a result of the two
  values' closeness.  We further see, that the NA62 constraints for
B6 and B8 exclude WC $\gtrsim 10^{-6} v^{-2}$, justifying the WC
choice in figure~\ref{fig:B5B8fixedWC}.  Overall, for benchmarks
B5--B8, the future far detectors will access new parameter space for
larger HNL masses, while DUNE shows
  much more promising sensitivities even for $m_N$ as low as 0.25 GeV.

\section{Summary}\label{sec:summary}

In this work, we have studied the potential of present and future far-detector experiments at the LHC and the beam-dump-type experiment DUNE, for probing long-lived heavy neutral leptons (HNLs) of Majorana nature produced from rare kaon
decays, in the theoretical framework of low-energy effective field
theory extended with sterile neutrinos ($N_R$LEFT).  We have focused
on dimension-6 effective operators consisting of a pair of quarks
together with a charged lepton and an HNL, or a pair of HNLs.  Besides
the effective operators, we also take into account the minimal mixing
parameter between the HNLs and the standard-model active neutrinos.
For simplicity, we assume in this work that there is only one
kinematically relevant HNL, that the HNL mixes with the electron
neutrino only, and that for the effective operators also only the
first-generation leptons are considered.

There are both lepton-number-conserving and lepton-number-violating operators; we have investigated both of them and elaborated on their differences.
We have computed the kaons' decay branching ratios into the HNLs with the considered effective operators, as a function of the HNL mass, the effective couplings, and the mixing parameter $U_{eN}$.
In addition, the decay rates of the HNLs in the $N_R$LEFT are calculated with care, including the interference between the EFT operators and the minimal mixing contributions.
We have further performed detailed Monte-Carlo simulations, in order to determine the acceptance of the LHC far detectors and the DUNE-ND for the long-lived HNLs.
The LHC experiments include ANUBIS, CODEX-b, FASER and FASER2, FACET, MATHUSLA, and MoeDAL-MAPP1 and MAPP2.
In particular, because of the long-lifetime nature of the kaons ($K^\pm$, $K_S$, and $K_L$), we cannot assume they decay essentially at the IPs; we have thus taken into account their decay positions in the simulation when we compute the decay probability of the HNLs in the detector fiducial volumes.
Moreover, for the various experiments, we have properly placed a cut-off position for each LHC far detector beyond which the kaons are vetoed.

For a series of benchmark scenarios classified by the number of HNLs in the operators as well as the Lorentz structure of the operators, we have obtained numerical results.
Besides the projection for the considered experiments, we have recast existing bounds on the HNLs in the minimal mixing scenario into those on the HNLs in the considered EFT benchmarks.
We find that for the pair-$N_R$ scenarios, the existing bounds from NA62 are already so strong that it has excluded all the parameter space that could be probed by future LHC far-detector experiments, but DUNE can still be sensitive to new parameter space in the case of scalar-type operators (benchmarks B2.1 and B2.2).
On the other hand, for the single-$N_R$ benchmarks, the studied future experiments are sensitive to regions of parameter space currently unexcluded.
Particularly, for benchmarks B3 and B4, the projected limits on the effective couplings can be orders of magnitude stronger than the existing bounds.
In all these benchmarks, we find the projected sensitivities for DUNE-ND are stronger than those for the LHC far detectors by various degrees in different benchmarks, mainly in virtue of the much larger production rates of the kaons at DUNE.

To summarize, our findings in this work show that for long-lived HNLs in the EFT framework produced from kaons, the DUNE-ND is expected to have sensitivities much more promising than the present and future LHC far detectors.
	Nevertheless, the LHC far detectors can probe unexcluded parameter space in some scenarios, motivating their construction and operation during the HL phase of the LHC.

\section*{Acknowledgements}


We thank Juan Carlos Helo and Felix Kling for useful discussions, and thank Giovanna Cottin for contributions in the early stage of the project.
This work is supported by the Spanish grants PID2020-113775GB-I00 (AEI/10.13039/ 501100011033) and CIPROM/2021/054 (Generalitat Valenciana).
R.B. acknowledges financial support from the Generalitat Valenciana (grant ACIF/2021/052).

\appendix
\section{Kaon decays into HNLs}
\label{app:kaondecays}
Throughout this work, we assume $N$ to be a Majorana particle. 
We neglect $|\varepsilon| \ll 1$ responsible for indirect $CP$ violation in the neutral kaon system, such that $K_S$ and $K_L$ coincide with 
$K_1$ and $K_2$ given by eq.~\eqref{eq:K1K2}.
Since the latter are superpositions of flavor eigenstates 
and the effective operators we consider are written in the flavor basis,
it proves convenient to define the following combinations of the Wilson coefficients:
\begin{align}
 a^V_{ij} &\equiv c_{dN,ij}^{V,RR} - c_{dN,ij}^{V,LR}
 \qquad \text{and} \qquad
 a^S_{ij} \equiv c_{dN,ij}^{S,RR} - c_{dN,ij}^{S,LR}\,, \\
 b^V_{ij} &\equiv c_{dN,ij}^{V,RR} + c_{dN,ij}^{V,LR}
 \qquad \text{and} \qquad
 b^S_{ij} \equiv c_{dN,ij}^{S,RR} + c_{dN,ij}^{S,LR}\,.
\end{align}
In addition, for the single-$N_R$ operators with a charged lepton, we define
\begin{align}
 c^V_{ij} &\equiv c_{udeN,ij}^{V,RR} - c_{udeN,ij}^{V,LR}
 \qquad \text{and} \qquad
 c^S_{ij} \equiv c_{udeN,ij}^{S,RR} - c_{udeN,ij}^{S,LR}\,, \\
 d^V_{ij} &\equiv c_{udeN,ij}^{V,RL} - c_{udeN,ij}^{V,LL}
 \qquad \text{and} \qquad
 d^S_{ij} \equiv c_{udeN,ij}^{S,RL} - c_{udeN,ij}^{S,LL}\,, \\
 g^V_{ij} &\equiv c_{udeN,ij}^{V,RR} + c_{udeN,ij}^{V,LR} 
 \qquad \text{and} \qquad
 g^S_{ij} \equiv c_{udeN,ij}^{S,RR} + c_{udeN,ij}^{S,LR}\,, \\
 h^V_{ij} &\equiv c_{udeN,ij}^{V,RL} + c_{udeN,ij}^{V,LL} 
 \qquad \text{and} \qquad
 h^S_{ij} \equiv c_{udeN,ij}^{S,RL} + c_{udeN,ij}^{S,LL}\,.
\end{align}
%

\subsection{Form factors}

The non-zero hadronic matrix elements entering the computations of kaon decay amplitudes are given by
\begin{align}
\langle 0 | \bar{s} \gamma^\mu \gamma_5 d| K^0 \rangle & = i f_K p_K^\mu \,,\\
\langle 0 | \bar{s} \gamma_5 d | K^0 \rangle & = i \frac{m_K^2}{m_d + m_s} f_K \equiv i f_K^S \,,
\end{align}
for leptonic kaon decays. Here $m_K = 493.6$~MeV is the neutral kaon mass, 
and $f_K = 155.7$~MeV~\cite{FLAG:2021npn}, and $f_K^S$ are the kaon decay constants. 
For $K^0 \rightarrow \pi^-$ transitions, the relevant matrix elements read
\begin{align}
\langle \pi^- | \bar{s} \gamma^\mu  u| K^0 \rangle & = f_+(q^2) P^\mu + \left(f_0(q^2) - f_+(q^2)\right) \frac{m_K^2 - m_\pi^2}{q^2} q^\mu \,,\\[3pt]
\langle \pi^- | \bar{s} u | K^0 \rangle & =  \frac{m_K^2-m_\pi^2}{m_u - m_s} f_0(q^2)\equiv f_S(q^2) \,,\\[4pt]
\langle \pi^- | \bar{s} \sigma^{\mu\nu} u | K^0 \rangle & =  \frac{i}{m_K} \left(p_K^\mu p_\pi^\nu - p_K^\nu p_\pi^\mu \right) B_T(q^2) \,,
\end{align}
where $P= p_K + p_\pi$ is the sum of the kaon and pion 4-momenta, and $q = p_K- p_\pi$ is their difference.
The three form factors ($f_+$, $f_0$, $B_T$) can be parameterized in terms of $q^2$ as follows (see Ref.~\cite{Falkowski:2021bkq} for more detail)
\begin{align}
 f_+(q^2) &= f_+(0) + \Lambda_+ \frac{q^2}{m_\pi^2}\,, 
 \label{eq:fp} \\
 f_0(q^2) &= f_+(0) + \left(\log C - G(0) \right) \frac{m_\pi^2}{m_K^2 - m_\pi^2} \frac{q^2}{m_\pi^2}\,, 
 \label{eq:f0} \\
 B_T(q^2) &= B_T(0) \left(1 - s_T^{K\pi} q^2\right)\,, 
 \label{eq:BT}
\end{align}
and the numerical values of the parameters entering these expressions are reported in table~\ref{tab:FFs}.
\begin{table}[t]
\renewcommand{\arraystretch}{1.3}
\centering
 \begin{tabular}[t]{|l|c|c|c|c|c|c|}
 \hline
 Parameter & $f_+(0)$ & $\Lambda_+$ & $\log C$ & $G(0)$ & $B_T(0)/f_+(0)$ & $s_T^{K\pi}~[\text{GeV}^{-2}]$ \\
 \hline
 Central value & 0.9706 & 0.02422 & 0.1198 & 0.0398 & 0.68 & 1.10 \\
 \hline
 \end{tabular}
 \caption{Parameters entering the form factors $f_+$, $f_0$ and $B_T$ in eqs.~\eqref{eq:fp}--\eqref{eq:BT}.}
 \label{tab:FFs}
\end{table}
%
\subsection{Two-body decays}
The partial decay widths of $K_{S/L} \to N N$ mediated by 
the pair-$N_R$ operators $\mathcal{O}_{dN}$
given in table~\ref{tab:opsNN} read: 
\begin{align}
 \Gamma(K_S \to N N) &= \frac{m_K}{32\pi}  
 \sqrt{1 - \frac{4m_N^2}{m_K^2}} \,
 \bigg[4 f_K^2 m_N^2
 \left(\im a^V_{21}\right)^2 \nonumber \\
 & + \left(f_K^S\right)^2 \bigg\{
 \left[\im \left(a^S_{21} - a^S_{12}\right)\right]^2 
 + \left[\re \left(a^S_{21} - a^S_{12}\right)\right]^2
 \left(1 - \frac{4 m_N^2}{m_K^2}\right) 
 \bigg\} \nonumber \\
 &+4 f_K f_K^S m_N \im a^V_{21} \im\left(a^S_{21} - a^S_{12}\right)\bigg]\,, 
 \label{eq:GammaKSNN} \\
 \Gamma(K_L \to N N) &= \frac{m_K}{32\pi}
 \sqrt{1 - \frac{4m_N^2}{m_K^2}} \,
 \bigg[4 f_K^2 m_N^2
 \left(\re a^V_{21}\right)^2 \nonumber \\
 & + \left(f_K^S\right)^2 \bigg\{
 \left[\re \left(a^S_{21} + a^S_{12}\right)\right]^2 
 + \left[\im \left(a^S_{21} + a^S_{12}\right)\right]^2
 \left(1 - \frac{4 m_N^2}{m_K^2}\right) 
 \bigg\} \nonumber \\
 &+4 f_K f_K^S m_N \re a^V_{21} \re\left(a^S_{21} + a^S_{12}\right)\bigg]\,.
 \label{eq:GammaKLNN} 
\end{align}
In this computation, we have neglected the contribution 
from active-heavy mixing 
for two reasons: (i) the amplitude for $K_{S/L}\to NN$ is proportional to $U_{\ell N}^2$ and $|U_{\ell N}|^2 \ll 1$,
and (ii) flavor-changing neutral currents are further suppressed, 
since in the SM they do not occur at tree level. 

The single-$N_R$ operators with a charged lepton 
summarized in table~\ref{tab:opseN} as well as the standard 
active-heavy neutrino mixing trigger $K^- \to \ell^- N$. 
For the corresponding decay width, we find
\begin{align}
 \Gamma(K^- \to \ell^- N) &= \frac{\lambda^{1/2}\left(m_K^2,m_\ell^2,m_N^2\right)}{64\pi m_K^3} 
\bigg\lbrace
   f_K^2 \left( \left|c^V_{12}\right|^2 + \left|d^V_{12}\right|^2 + 
   |c_{\text{mix}}|^2 - 2\re\left(d^V_{12} c_{\text{mix}}^\ast\right) \right) \nonumber \\
  &\times \left[m_K^2 \left(m_\ell^2+m_N^2\right) - \left(m_\ell^2-m_N^2\right)^2 \right] \nonumber \\
  &+\left(f_K^S\right)^2 \left( \left|c^S_{12}\right|^2 + \left|d^S_{12}\right|^2 \right) 
  \left[m_K^2-m_\ell^2-m_N^2\right] \nonumber \\
  &+ 2 f_K f_K^S \re\left(c^V_{12} c^{S\ast}_{12} + d^V_{12} d^{S\ast}_{12} 
  - d^{S \ast}_{12} c_{\text{mix}}\right) 
  m_\ell \left[m_K^2 - m_\ell^2 + m_N^2\right] \nonumber \\
  &-4 f_K^2 \re\left(c^V_{12} d^{V\ast}_{12} - c^V_{12} c_{\text{mix}}^\ast\right) m_K^2 m_\ell\, m_N
  -4 \left(f_K^S\right)^2 \re\left(c^S_{12} d^{S\ast}_{12}\right) m_\ell\, m_N \nonumber \\
  &-2 f_K f_K^S \re\left(c^V_{12} d^{S\ast}_{12} + c^S_{12} d^{V\ast}_{12} 
  - c^{S}_{12} c_{\text{mix}}^\ast\right) m_N \left[m_K^2 + m_\ell^2 - m_N^2\right]
 \bigg\rbrace\,,
 \label{eq:GammaKeN}
\end{align}
where $\lambda(x,y,z) = x^2+y^2+z^2-2xy-2xz-2yz$.
This result complements eq.~(52) of Ref.~\cite{DeVries:2020jbs}, 
where only one LNV operator 
($C_\mathrm{VLL}^{(6)}$ in their notation) 
arising from the $N_R$SMEFT at $d=6$ has been included. 
In our notation, it means 
$d^V_{12} = - c_{udeN,12}^{V,LL}$ and $d^S_{12} = 0$.
We find agreement with eq.~(52) of Ref.~\cite{DeVries:2020jbs} 
in this limit.

\subsection{Three-body decays}
We rely on the isospin symmetry and express the results 
in terms of the form factors $f_+$, $f_0$, $f_S$ and $B_T$ 
for $\langle\pi^-| \overline{s}\, \Gamma\, u| K^0\rangle$ 
given in eqs.~\eqref{eq:fp}--\eqref{eq:BT}. 
By the isospin symmetry, we have
\begin{align}
 \langle\pi^0| \overline{s}\, \Gamma\, d| K^0\rangle &= - \frac{1}{\sqrt2}\langle\pi^-| \overline{s}\, \Gamma\, u| K^0\rangle\,, 
  \label{eq:K0pi0ME} \\
 \langle\pi^+| \overline{s}\, \Gamma\, d| K^+\rangle &= 
 \langle\pi^-| \overline{s}\, \Gamma\, u| K^0\rangle\,. 
 \label{eq:KppipME}
\end{align}

The amplitude for $K_{S/L} \to \pi^0 N N$ is given by
\begin{align}
 \mathcal{M}(K_{S/L}\to \pi^0 N N) &= \frac{1}{4} \left(b^V_{21} \pm b^{V\ast}_{21}\right)
 \left[f_{+} P_{\mu}  + \left(f_{0} - f_{+}\right)\frac{m_K^2-m_\pi^2}{q^2} q_{\mu} \right] 
 \left[\overline{u_N} \gamma^{\mu} \gamma_5 v_{N'}\right] \nonumber \\
  & + \frac{f_S}{2} \left(b^{S}_{21} \mp b^{S}_{12}\right) 
  \left[\overline{u_N} P_R v_{N'}\right] 
  + \frac{f_S}{2} \left(b^{S}_{12} \mp b^{S}_{21}\right)^\ast 
  \left[\overline{u_N} P_L v_{N'}\right]\,.
\end{align}
The squared amplitudes summed over spins then read:
\begin{align}
 \sum_{\mathrm{spins}} \left|\mathcal{M}(K_S \to \pi^0 N N)\right|^2
 &= 2 \left(\re b^V_{21}\right)^2 
 \bigg\{-4f_+^2 (p\cdot p_N)^2 + 2f_+^2 (a+ m_K^2-m_\pi^2)(p \cdot p_N) \nonumber \\ 
 & - f_+^2  \left[  a(m_K^2 + m_N^2) -2 m_N^2 (m_K^2 + m_\pi^2) \right] 
 + \frac{(f_0^2-f_+^2)}{a} m_N^2 (m_K^2-m_\pi^2)^2 \bigg\} \nonumber \\
 & +\frac{ f_S^2}{2}\, \bigg\{\left[\re\left(b^{S}_{21}-b^{S}_{12}\right)\right]^2 a 
 + \left[\im\left(b^S_{21}-b^S_{12}\right)\right]^2 \left(a - 4m_N^2\right)
 \bigg\} \nonumber \\
 &+2 f_0 f_S\, m_N \left(m_K^2 - m_\pi^2\right) \re b^V_{21} \re\left(b^S_{21}-b^S_{12}\right)\,, 
 \label{eq:GammaKSpi0NN} \\[0.2cm]
  \sum_{\mathrm{spins}} \left|\mathcal{M}(K_L \to \pi^0 N N)\right|^2
 &= 2 \left(\im b^V_{21}\right)^2 
 \bigg\{-4f_+^2 (p\cdot p_N)^2 + 2f_+^2 (a+ m_K^2-m_\pi^2)(p \cdot p_N) \nonumber \\ 
 & - f_+^2  \left[  a(m_K^2 + m_N^2) -2 m_N^2 (m_K^2 + m_\pi^2) \right] 
 + \frac{(f_0^2-f_+^2)}{a} m_N^2 (m_K^2-m_\pi^2)^2 \bigg\} \nonumber \\
 & + \frac{f_S^2}{2}\, \bigg\{\left[\im\left(b^{S}_{21}+b^{S}_{12}\right)\right]^2 a 
 + \left[\re\left(b^S_{21}+b^S_{12}\right)\right]^2 \left(a - 4m_N^2\right)
 \bigg\} \nonumber \\
 &+2 f_0 f_S\, m_N \left(m_K^2 - m_\pi^2\right) \im b^V_{21} \im\left(b^S_{21}+b^S_{12}\right)\,.
 \label{eq:GammaKLpi0NN}
\end{align}

For $K^+ \to \pi^+ N N$, we find:
\begin{align}
 \mathcal{M}(K^+ \to \pi^+ N N) &= \frac{b^V_{21}}{2}
 \left[f_{+} P_{\mu}  + \left(f_{0} - f_{+}\right)\frac{M^2-m^2}{q^2} q_{\mu} \right] 
 \left[\overline{u_N} \gamma^{\mu} \gamma_5 v_{N'}\right] \nonumber \\
 &+ f_S\, b^S_{21} \left[\overline{u_N} P_R v_{N'}\right] 
 + f_S\, b^{S\ast}_{12} \left[\overline{u_N} P_L v_{N'}\right]\,,
\end{align}
and
\begin{align}
 \sum_{\mathrm{spins}} \left|\mathcal{M}(K^+ \to \pi^+ N N)\right|^2
 &= 2 \left|b^V_{21}\right|^2 
 \bigg\{-4f_+^2 (p\cdot p_N)^2 + 2f_+^2 (a+ m_K^2-m_\pi^2)(p \cdot p_N) \nonumber \\ 
 & - f_+^2  \left[  a(m_K^2 + m_N^2) -2 m_N^2 (m_K^2 + m_\pi^2) \right] 
 + \frac{(f_0^2-f_+^2)}{a} m_N^2 (m_K^2-m_\pi^2)^2 \bigg\} \nonumber \\
 & + f_S^2\, \bigg\{\left(\left|b^S_{21}\right|^2 + \left|b^S_{12}\right|^2\right) \left(a - 2m_N^2\right)
 - 4m_N^2 \re \left(b^S_{21} b^S_{12}\right)
 \bigg\} \nonumber \\
 &+2 f_0 f_S\, m_N \left(m_K^2 - m_\pi^2\right) \re \left[b^V_{21} \left(b^{S\ast}_{21}-b^S_{12}\right)\right]\,.
\end{align}

Turning to the single-$N_R$ operators with a charged lepton, 
we have 
\begin{align}
 \mathcal{M}(K^- \to \pi^0 \ell^- N) &= \frac{1}{2\sqrt{2}} \bigg\lbrace g^V_{12}\left[\overline{u_\ell} \gamma^{\mu} P_R v_{N}\right] 
 +(h^V_{12} + c_{\text{mix}}) \left[\overline{u_\ell} \gamma^{\mu} P_L v_{N}\right]\bigg\rbrace  \nonumber  \\ 
& \times \left[f_{+} P_{\mu}  + \left(f_{0} - f_{+}\right)\frac{m_K^2-m_\pi^2}{q^2} q_{\mu} \right] \nonumber \\
 &+\frac{f_S}{2\sqrt{2}}
 \bigg\lbrace g^S_{12} \left[\overline{u_\ell} P_R v_N\right] 
 + h^S_{12} \left[\overline{u_\ell} P_L v_N\right] \bigg\rbrace \nonumber \\
 &+\frac{i}{2\sqrt{2}} \frac{B_T}{m_K} 
 \bigg\lbrace
 \left[p^{\mu} p^{\prime\nu} - p^{\nu} p^{\prime\mu} 
 + i\, \epsilon_{\mu\nu\alpha\beta}\, p^\alpha p^{\prime\beta}\right] 
 c_{udeN,12}^{T,RR} \left[\overline{u_\ell} \sigma^{\mu\nu} P_R v_N\right] \nonumber \\
&
+\left[p^{\mu} p^{\prime\nu} - p^{\nu} p^{\prime\mu} 
 - i\, \epsilon_{\mu\nu\alpha\beta}\, p^\alpha p^{\prime\beta}\right] 
c_{udeN,12}^{T,LL} \left[\overline{u_\ell} \sigma^{\mu\nu} P_L v_N\right] 
 \bigg\rbrace\,.
 \label{eq:MKmpi0eN}
\end{align}
%
Finally, up to an unphysical sign,
\begin{equation}
 \mathcal{M}(K_{S/L} \to \pi^+ \ell^- N) = \mathcal{M}(K^- \to \pi^0 \ell^- N)\,.
\end{equation} 
We do not provide the full expression for the modulus square of amplitude~\eqref{eq:MKmpi0eN} summed over spins, since 
it is rather cumbersome. 
However, it is straightforward to obtain it, 
especially, assuming only one operator (either V, S, or T) at a time. 
In the limit of zero mixing ($c_{\text{mix}}=0$), as in the case of two-body decay in eq.~\eqref{eq:GammaKeN}, 
the LNV single-$N_R$ operators (switched on one at a time) 
lead to the same results as their LNC counterparts. 
Finally, the three-body decay widths are computed 
following the procedure explained in Refs.~\cite{DeVries:2020jbs,Beltran:2022ast}.

\section{HNL decays}
\label{app:HNLdecays}
Effective operators in $N_R$LEFT not only enhance the HNL production but also trigger their decay. Assuming only one generation of $N_R$, the pair-$N_R$ operators in table~\ref{tab:opsNN} cannot make the HNL decay, whereas the single-$N_R$ operators in table~\ref{tab:opseN} do contribute to it. In this appendix we provide the partial decay width of $N \rightarrow \ell^- \pi^+$, which is the only kinematically allowed channel if HNLs are produced in kaon decays and if $m_N > m_\pi + m_\ell$. In the computation we take into account the contribution from (i) the $\mathcal{O}_{udeN}$ operators in table~\ref{tab:opseN}, (ii) eq.~\eqref{eq:Lmix}, \textit{i.e.} the standard mixing to active neutrinos, and (iii) the interference terms between (i) and (ii). The final expression reads
\begin{align}
 \Gamma(N \to \ell^- \pi^+) &= \frac{\lambda^{1/2}\left(m_N^2,m_\ell^2,m_\pi^2\right)}{128\pi m_N^3}
  \bigg\lbrace  f_\pi^2 \left( \left|c^V_{11}\right|^2 + \left|d^V_{11}\right|^2 
  + |c_{\text{mix}}|^2  - 2\re\left(d^V_{11} c_{\text{mix}}^\ast\right)  \right) \nonumber \\
& \times \left[\left(m_\ell^2-m_N^2\right)^2  - m_\pi^2 \left(m_\ell^2+m_N^2\right) \right] \nonumber \\
  &+\left(f_\pi^S\right)^2 \left( \left|c^S_{11}\right|^2 + \left|d^S_{11}\right|^2 \right) 
  \left[m_\ell^2+m_N^2-m_\pi^2\right] \nonumber \\
  &- 2 f_\pi f_\pi^S \re\left(c^V_{11} c^{S\ast}_{11} + d^V_{11} d^{S\ast}_{11}  -  c_{\text{mix}} d^{S\ast}_{11}  \right) 
  m_\ell \left[m_\ell^2 - m_N^2- m_\pi^2 \right] \nonumber \\
  &+ 4 f_\pi^2 \re\left(c^V_{11} d^{V\ast}_{11} - c^V_{11} c^\ast_{\text{mix}} \right) m_\pi^2 m_\ell\, m_N
  +4 \left(f_\pi^S\right)^2 \re\left(c^S_{11} d^{S\ast}_{11}\right) m_\ell\, m_N \nonumber \\
  &-2 f_\pi f_\pi^S \re\left(c^V_{11} d^{S\ast}_{11} + c^S_{11} d^{V\ast}_{11} - c^S_{11} c^\ast_{\text{mix}} \right) m_N \left[m_\ell^2 - m_N^2 + m_\pi^2\right]
 \bigg\rbrace\,.
 \label{eq:GammaNlpi}
\end{align}
The previous result can be derived from the amplitude leading to eq.~\eqref{eq:GammaKeN} by substituting $K$ with $\pi$ and perfomring the interchange $p_\pi \leftrightarrow p_N$ along the computation. The decay constants are given by $f_\pi = 130.2$\,MeV~\cite{FLAG:2021npn} and $f_\pi^S=\frac{m_\pi^2}{m_u + m_d} f_\pi$. We have also used the notation introduced in appendix~\ref{app:kaondecays} for the WC, and $c_{\rm mix}$ is the coefficient defined in eq.~\eqref{eq:cmix}.

In the scenarios outlined in table~\ref{tab:benchmarks}, the decay of the HNL is governed by either the neutrino mixing parameter (scenarios B1--B2 and B5--B8) or one of the WC of the effective operators (scenarios B3 and B4). In cases where the former applies, eq.~\eqref{eq:GammaNlpi} reduces to the one in the minimal scenario \cite{Bondarenko:2018ptm}. Conversely, in the latter cases, eq.~\eqref{eq:GammaNlpi} can be simplified to
\begin{align}
 \Gamma(N \to \ell^- \pi^+) &= \frac{\lambda^{1/2}\left(m_N^2,m_\ell^2,m_\pi^2\right)}{128\pi m_N^3}
  \bigg\lbrace  f_\pi^2 \left( \left|c^V_{11}\right|^2 + \left|d^V_{11}\right|^2  \right) \left[\left(m_\ell^2-m_N^2\right)^2  - m_\pi^2 \left(m_\ell^2+m_N^2\right) \right] \nonumber \\
  &+\left(f_\pi^S\right)^2 \left( \left|c^S_{11}\right|^2 + \left|d^S_{11}\right|^2 \right) 
  \left[m_\ell^2+m_N^2-m_\pi^2\right] \nonumber \\
    &+ 4 f_\pi^2 \re\left(c^V_{11} d^{V\ast}_{11} \right) m_\pi^2 m_\ell\, m_N + 4 \left(f_\pi^S\right)^2 \re\left(c^S_{11} d^{S\ast}_{11}\right) m_\ell\, m_N \nonumber \\[4pt]
  &- 2 f_\pi f_\pi^S \re\left(c^V_{11} c^{S\ast}_{11} + d^V_{11} d^{S\ast}_{11} \right) 
  m_\ell \left[m_\ell^2 - m_N^2- m_\pi^2 \right] \nonumber \\
  &-2 f_\pi f_\pi^S \re\left(c^V_{11} d^{S\ast}_{11} + c^S_{11} d^{V\ast}_{11} \right) m_N \left[m_\ell^2 - m_N^2 + m_\pi^2\right]
 \bigg\rbrace\,.
 \label{eq:GammaNlpiWC}
\end{align}
It is worth mentioning that in scenarios B1--B2 and B5--B8, there are additional decay channels for $N$, such as the purely leptonic $\nu\nu\nu$ and $\nu\ell\ell$ channels, which enhance the HNL total decay width. The possible open channels depend on the HNL mass \cite{Bondarenko:2018ptm}. Meanwhile, in scenarios B3--B4, the total decay width becomes twice the result in eq.~\eqref{eq:GammaNlpiWC}, since the charge conjugated  channel is also open for Majorana $N$.

\bibliographystyle{JHEP}
\bibliography{hnl_kaon}

\end{document}